\documentclass[final,5p,times,twocolumn]{elsarticle}

\usepackage{algorithm2e}
\usepackage{amsmath}
\usepackage{amsmath,algpseudocode}\usepackage{algcompatible}
\usepackage{amssymb}
\usepackage{amsthm} 
\usepackage{color,soul}
\usepackage{csquotes}
\usepackage{epsfig}
\usepackage{float}
\usepackage{float}
\usepackage{graphicx}
\usepackage{latexsym}
\usepackage{lineno}
\usepackage{listings}
\usepackage{lscape}
\usepackage{mathtools}
\usepackage{siunitx}
\usepackage{tabularx,ragged2e,booktabs,caption,array,multirow,multicol}
\usepackage{tabularx}
\usepackage{subfig}
\usepackage{verbatim}
\usepackage{xcolor}
\usepackage{xspace}
\usepackage[normalem]{ulem}

\usepackage[export]{adjustbox}
\usepackage{multirow}

\newcommand{\mb}[1]{\mathbf{#1}}
 \newcommand{\bs}[1]{\boldsymbol{#1}}
 
\newlength\myindent
\journal{Environmental Modelling \& Software}

\begin{document}
\begin{frontmatter} 

\title{\textit{Bayesreef}: A Bayesian inference framework for modelling reef growth in response to environmental change and biological dynamics
}

\author[grg,ctds]{Jodie Pall } 
\author[ctds,grg]{Rohitash Chandra\corref{cor2}}
\ead{rohitash.chandra@sydney.edu.au}
\author[grg,ctds]{Danial Azam}
\author[grg]{Tristan Salles}
\author[grg]{Jody M. Webster}
\author[ctds]{Richard Scalzo} 
\author[smas,ctds]{Sally Cripps}
\cortext[cor2]{Corresponding author}
\address[grg]{Geocoastal Research Group, School of Geosciences, University of 
Sydney, NSW 2006\\Sydney, Australia}
\address[ctds]{Centre for Translational Data Science, University 
of Sydney, NSW 2006\\Sydney, Australia}
\address[smas]{School of Mathematics and Statistics, University 
of Sydney, NSW 2006\\Sydney, Australia}
\begin{abstract}  
  Estimating the impact of environmental processes on vertical reef development in geological time is a very challenging task.  pyReef-Core is a deterministic carbonate stratigraphic forward model designed to simulate the key biological and environmental processes that determine vertical reef accretion and assemblage changes in fossil reef drill cores.     We present a Bayesian framework called Bayesreef for the estimation and uncertainty quantification of parameters in pyReef-Core that represent environmental conditions affecting the growth of coral assemblages on geological timescales. We demonstrate the existence of multimodal posterior distributions and investigate the challenges of sampling using Markov chain Monte-Carlo (MCMC) methods, which includes parallel tempering MCMC.   We use synthetic reef-core to investigate fundamental issues and then apply the methodology to a selected reef-core from the Great Barrier Reef in Australia. The results  show that Bayesreef accurately estimates and provides uncertainty quantification of the selected  parameters that represent environment and ecological conditions in pyReef-Core.  Bayesreef provides insights into the complex posterior distributions of parameters in  pyReef-Core, which provides the groundwork for future research in this area.

\end{abstract}

\end{frontmatter}

\section{Introduction}
\label{intro}

 Developing data-driven models of reef evolution is challenging because the complexity of the process exceeds the amount of available data necessary.  Reef evolution is determined by the interaction between environmental factors such as water chemistry, light availability, sedimentation and hydrodynamic energy \cite{Veron08}.  The data for understanding and modelling reef evolution can be extracted from the geological record of reef-drilled cores, which is sparse, and also expensive to obtain; hence, limited work has been done in this area.

 \textit{pyReef-Core} \cite{Salles2018} is an example of a carbonate stratigraphic   forward model (SFM) for reef evolution that captures a number of important ecological dynamics in coral reef systems.  \textit{pyReef-Core} is the first modelling attempt to constrain hydrodynamic energy and sediment input exposure thresholds for coralgal assemblages on a geological timescale. It is a one-dimensional (1-D) model  that simulates the vertical (and not lateral, hence 1-D) coralgal growth patterns observed in a reef drill core. 
\textit{pyReef-Core} has several parameters representing external environmental factors which impact reef development. Examples of these factors include sea-level changes and the relationship between sediment input and depth. It also has parameters describing the response of coralgal assemblage growth to these environmental factors, such as water flow and parameters for internal population dynamics such as the Malthusian 
 parameter. Figure \ref{fig:workflow} shows the workflow of {\it pyReef-Core}, which shows vertical accumulation contributed by different coralgal assemblages over a timeframe.
  
Given limited and sparse data, we need to quantify uncertainty from different factors in the estimation of unknown parameters in stratigraphic forward models such as \textit{pyReef-Core}.   Moreover, geophysical and stratigraphic forward models rarely have a unique solution \cite{isakov1993uniqueness,charvin2009bayesian} which is also known as non-uniqueness \cite{BurgessPrince15a}. For example, different combinations of a range of environmental parameters such as water flow, temperature and population dynamics of the coral assemblages in  {\it pyReef-Core}  may give rise to the same simulated reef-core stratigraphy.   Stratigraphic forward models produce a set of solutions that represent multiple and competing hypotheses regarding geological system evolution \cite{cross1999construction,heller1993stratigraphic}. However, the explicit temporal and depth structure simulated by \textit{pyReef-Core} presents an opportunity to restrict the number of possible solutions and thus reduce uncertainty.

   Bayesian inference provides a rigorous methodology for estimation and uncertainty quantification of unknown parameters in a given model by incorporating information from multiple sources \cite{robert2009}. The information from prior research,  expert opinion, and knowledge regarding the nature of specific physical processes can be incorporated via a set of prior beliefs, also known as \textit{priors}. \textcolor{black}{Moreover, information from observed data is used to update these prior beliefs via the likelihood function.}   In the case of environmental modelling, Bayesian inference for uncertainty quantification has been deployed for a number of problems \cite{Anthony2012,hassan2009using,raje2012bayesian,Jens2007UncertainEnvi}.   

   We present a novel probabilistic framework for the estimation and uncertainty quantification of environmental processes and factors which impact the depth and temporal distribution of communities of corals and coralline algae (coralgal assemblages) found in fossil reef drill cores. The framework is called \textit{Bayesreef} which employs  Bayesian inference using Markov Chain Monte Carlo (MCMC) sampling to estimate the unknown parameters of {\it pyReef-Core}    Although we chose  \textit{pyReef-Core} to demonstrate the idea,  the framework is general, and can be adapted, for other stratigraphic forward models. The goal of {\it Bayesreef} is to provide  estimation and uncertainty quantification for complex processes with sparse data which  presents  four significant contributions to the literature. 

 First, we transform a selected deterministic stratigraphic forward model (\textit{pyReef-Core} )    into a probabilistic framework known as  \textit{Bayesreef},   where the parameters are sampled via MCMC and represented using a probability distribution. Bayesreef employs a multinomial likelihood using the data from drilled reef-core featuring the coralgal assemblages over a timeframe.   The expert opinion and the results of previous studies are incorporated into the prior, while knowledge of the physical processes from the {\it pyReef-Core} is connected to the observed assemblage in the reef core via the multinomial likelihood.  

Second, we use  \textit{Bayesreef}  to constrain the number of solutions that represent the unique palaeo-environmental history of the reef core by incorporating two different types of data representation, where one features the temporal structure and the other the depth structure of the reef core. By depth structure, we refer to the thickness and type of   coralgal assemblage at varying depths.  By temporal structure,  we refer to  the thickness and type of sediment and/or coralgal material laid down at varying points in time. We use this feature of {\it pyReef-Core} to show how incorporating the time series structure constrains the number of possible outcomes. 

\textcolor{black}{ Third, we demonstrate the effectiveness of Bayesreef with two sampling methods, that includes single-chain MCMC and parallel tempering MCMC, for synthetic and real reef-core drilled from a selected location in the   \textit{ Great Barrier Reef}. }

Fourth, we make the methodology and its implementation available for other researchers as a software tool \footnote{Bayesreef: https://github.com/intelligentEarth/BayesReef}, which can be used to make inference regarding the factors affecting reef evolution.

 The rest of the paper is outlined as follows: Section 2 provides background and related work, while Section 3 presents the   methodology and techniques used including the multinomial likelihood function. Section 4 presents experiments and results. Section  5 provides discussion, and Section 6 concludes the paper with a discussion of future research.

\section{Background}

\subsection{Coral reef evolution}

The ability of corals to vigorously grow and build reef structures is dependent upon favourable environmental conditions \cite{done11ECR}. Three related environmental factors examined in \textit{pyReef-Core} are vital in influencing coral reef evolution on multi-decadal to centennial timescales. They are water depth (accommodation), hydrodynamic energy, and autochthonous (reef-derived) sediment input.

The accommodation is the vertical space in the water column above the substrate within which corals can grow and affects hydrodynamic energy and sediment flux. Hydrodynamic (wave and water flow) energy decreases with depth, such that corals growing in shallower water experience increased hydrodynamic energy \cite{Montaggioni05}. At the organism level, currents, water flow and oscillatory motion induced by waves are critical in modulating physiological processes in coral and thus influence coral growth rates \cite{Falteretal04,LoweFalter15}.  Similarly, fluxes of reef-derived carbonate sediments typically increase with depth as they are less disturbed by currents and settle on corals \cite{chappell1980coral}. The sediment input inhibits coral reef growth and even causes mortality via turbidity, reducing light and the ability of corals to meet energy requirements via photosynthesis \cite{Erftemeijeretal12, Rogers90}, and via smothering and abrasion \cite{sanders2005scleractinian}.

Over geological timescales, these environmental disturbances are essential determinants of the composition of coralgal assemblages and their spatial distribution in specific environmental niches across the reef and with depth
\cite{Montaggionietal97,HongoKayanne10,Kench11EMCR,camoin2015coral}. However, inferring the long-term trend of these processes from a geological perspective is difficult given missing data in the reef drill cores, which are an incomplete record of reef development through time. 

\begin{figure*}[ht!]
\centering
\includegraphics[width=180mm]{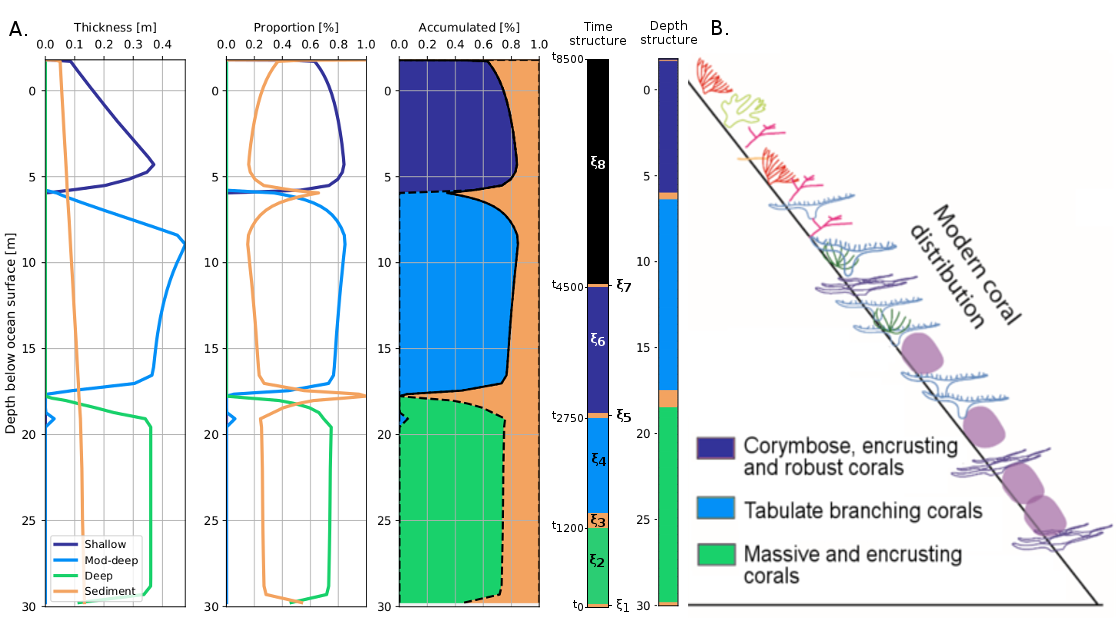}
\caption{(A)~\textit{pyReef-Core} output of the synthetic 30-metre-long reef drill core (ground truth) representing a catch-up growth sequence. The first column of \textit{pyReef-Core} output represents the amount of vertical accumulation contributed by the respective assemblages or by sediment deposition within the 50-year time intervals. The second and third columns represent the vertical accumulation of each assemblage as a proportion of each depth interval. Where coral growth is absent, the carbonate sand dominates according to \textit{pyReef-Core} model. The fourth column represents the amount of vertical accumulation in 50-year increments throughout the simulation. The fifth column represents the simulated drill core displaying the dominant assemblage (or sediment deposition) at each depth interval. (B)~A schematic of modern coral zonation with depth on a reef representing a shallowing-upward growth strategy, displaying associated assemblage compositions and transitions (adapted from Dechnik \cite{Dechnik16}).}
\label{fig:synthcore}
\end{figure*}

The palaeo-environmental  
analyses of drill cores have informed a prevailing theory of reef evolution in response to Holocene sea-level changes \cite{NeumannMacintyre85,Daviesetal85}. The shallowing-upward sequences are found in Holocene reef drill cores globally \cite{Braithwaiteetal04,Grigg98,HongoKayanne10}. An example of a drill core appears in Figure~\ref{fig:synthcore}.  The lower parts of drill cores are characterised by deep-water ($\sim$20-30~m), massive and branching corals, and bioclastic sediments. These are replaced  in shallower palaeo-water depths ($<$6~m) by robust-branching corals, representing high-energy coralgal assemblages \cite{Cabiochetal99,Montaggionietal97} (Figure~\ref{fig:synthcore}). The shallowing-upward sequences have been proposed to express a 'catch-up' vertical reef growth strategy, as coralgal vertical accumulation was not initially fast enough to keep up with rapid sea level rise, but caught up once sea level stabilised (Figure~\ref{fig:synthcore}).

\subsection{\textit{pyReef-Core}}

\begin{figure*}[htb!]
\centering
\noindent\includegraphics[width=40pc]{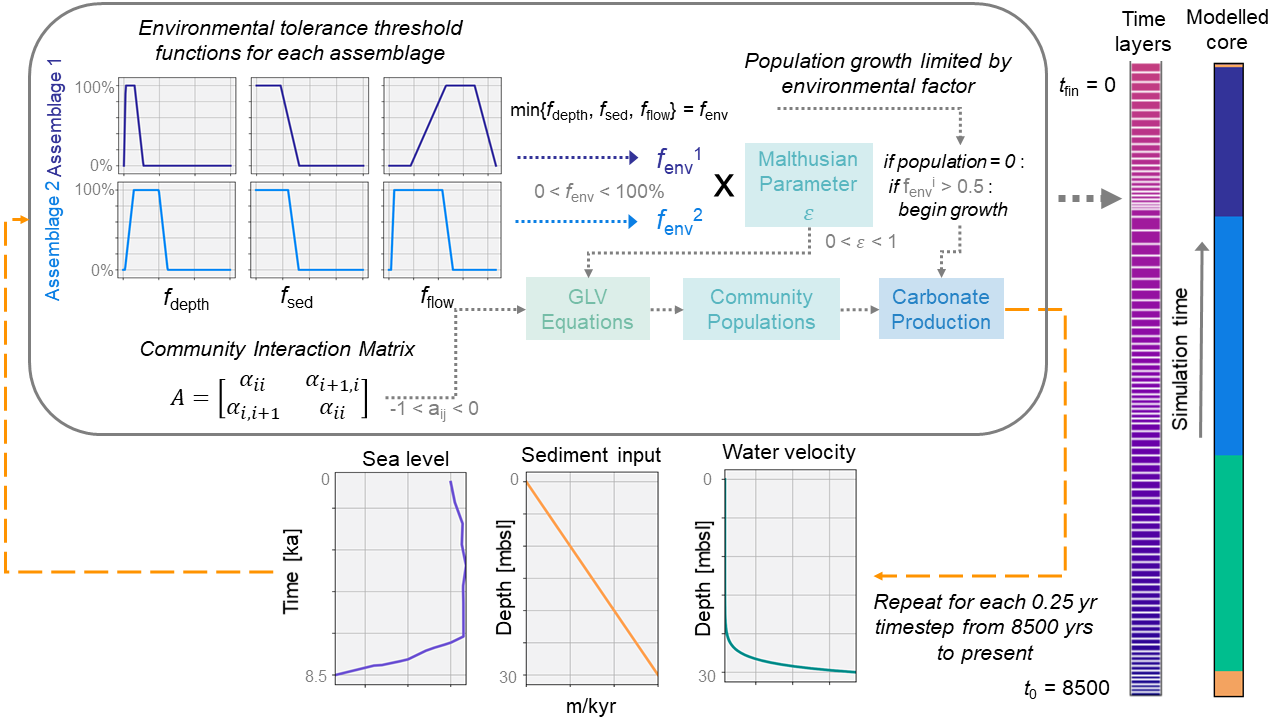}
\caption{Schematic of the \textit{pyReef-Core} workflow (left) and of the resulting simulated core (right). First boundary conditions for sea-level, sediment input and flow velocity are given. The bottom panel of figures describes how these quantities vary with time and depth. The environmental tolerance thresholds (upper left) establish the rate at which an assemblage grows vertically under the boundary conditions as a proportion of maximum growth. The environment factor ($f_{env}$) determines the percentage of the maximum growth rate that an assemblage can achieve. The environment factor 'turns on' growth when a minimum threshold is reached, determining the optimal conditions for initial growth. The environment factor is scaled by the Malthusian parameter, which is in turn used as input in the population dynamics (GLV) equations to determine assemblage populations. Larger assemblage population contribute to a faster rate of vertical accretion or carbonate production. At the end of the time step, the boundary conditions are updated, and the process is repeated (Salles et al. \cite{Salles2018}).}
\label{fig:workflow}
\end{figure*}

\textit{pyReef-Core} represents one of the first attempts to incorporate coral ecological dynamics into carbonate system modelling \citep{Salles2018}. The tool simulates the interaction of the main biological and physical reef-building processes, including hydrodynamic energy (flow velocity), sediment input and the ecological interactions between different coralgal assemblages (Figure \ref{fig:workflow}).

The physical process encoded in {\it pyReef-Core} provides a basis for constraining essential biological and physical processes to investigate the key drivers of a reef's evolution and thus forecast how they are affected by mid and long-term environmental changes.  Some of the input parameters to  {\it pyReef-Core} are observable, such as the relative sea-level history, depth-dependent rate of sediment input and depth-dependent water velocity (Figure~\ref{fig:boundary}). Other {\it pyReef-Core} parameters are not observed and unknown, and hence they need to be estimated. The unknown parameters include  those governing the population dynamics, the environmental threshold functions such as the intrinsic rate of growth/decline of coralgal assemblage populations, and parameters defining competitive dynamics between assemblages such as the assemblage interaction matrix (AIM). We incorporate the competitive coralgal assemblage interactions in \textit{pyReef-Core}  via the generalised Lotka-Volterra system of equations \cite{dafermos1970} (GLV), also known as predator-prey equations.

The environmental threshold function controls the rate of vertical accumulation for different coralgal assemblages which can be enhanced or limited by environmental factors (Figures~\ref{fig:workflow} and \ref{fig:thresholds}). The environmental threshold functions in \textit{pyReef-Core} measure the assemblage sensitivity to sediment input exposure and exposure to the hydrodynamic energy. In \textit{pyReef-Core}, four values define each exposure threshold, where the outer two values indicate the absolute minimum and maximum values of the environmental stress which are known to be tolerable to an assemblage and beyond which, the effects of exposure are lethal \cite{Erftemeijeretal12}. The remaining two values within the minimum and maximum bounds indicate where flow velocity or sediment input begins to restrict growth \cite{Salles2018}. During each time step, flow velocity and sediment input intersect different points of the threshold curves for each assemblage. The environmental factor, $F_{env}$, that limits growth during each time step is found by taking the minimum of all threshold functions (i.e. $f_{depth}$, $f_{sed}$ and $f_{flow}$; Figure \ref{fig:thresholds}). We multiply the environment factor by the maximum vertical accretion rate for each assemblage to limit growth according to environmental exposure (Figure \ref{fig:workflow}). 

While there are clear theoretical relationships between the duration and rates of sedimentation and flow velocity on coral mortality \cite{Erftemeijeretal12, baldock2014impact}, determining these thresholds quantitatively as inputs into \textit{pyReef-Core} remains challenging to estimate. In the next section, we use a data fusion approach to employ the existing state of knowledge within a Bayesian framework to estimate the unknown parameters.

\subsection{Bayesian inference}

It is common to denote unknown parameters generically by $\theta$; if there is only one parameter, or by $\bs\theta$ given in a vector of parameters. In the Bayesian paradigm, prior belief about $\theta$ is updated from information featured in an observed data point via the likelihood, and inference about $\theta$ proceeds via the posterior distribution. The relationship between these three quantities; the prior, the likelihood and the posterior is given by Bayes theorem
\begin{equation}
P(\bs \theta|\mathcal{D})=\frac{P(\mathcal{D}|\bs \theta)P(\bs \theta)}{P(\mathcal{D})}
\label{eqn_Bayes}
\end{equation}
where $\mathcal{D}$ is the data, $P(\mathcal{D}|\bs \theta)$ is the likelihood, $P(\bs \theta)$ is the prior and $P(\bs \theta|\mathcal{D})$ is the posterior. Unfortunately, this posterior distribution is rarely available in closed form which is particularly true for nonlinear inverse problems in geophysical models like \textit{pyReef-Core}, where no analytical expression for the forward relation between data and model parameters is available \citep{mosegaard2002monte}.  In situations such as these, we use MCMC sampling methods to approximate the posterior \cite{gelfand90} which involves proposing draws of the quantity of interest from some proposal distribution, and accepting them with a probability ensuring that the Markov chain is reversible \cite{hastings1970monte,metropolis1953equation}.

The applications of Bayesian inference via MCMC sampling methods have been well established in areas of Earth and environmental sciences such as applications to modelling geochronological ages \cite{jasra2006bayesian}, modelling the effect of climate changes in land surface hydrology \cite{raje2012bayesian}, inferring sea-level and sediment supply from the stratigraphic record \cite{Charvinetal09} and inferring groundwater contamination sources  \cite{wang2013characterization}. 
However, to our knowledge, no work has used Bayesian inference for reef modelling, despite evidence of their usefulness when handling models with complex, interrelating parameters \cite{brown2017tracing,franco2016bayesian}. However, there exists work on Bayesian belief networks to assess the rate of changes in coral reef ecosystems \cite{FRANCO2016} which further motivates the use of Bayesian methods for reef modelling.

  \textcolor{black}{Multimodal and discontinuous posterior distributions pose enormous challenges for MCMC methods \cite{gallagher2009markov}.  In the \textit{pyReef-Core} model, gradient information is not available which limits the use of MCMC sampling with gradient-based proposal distributions such as Hamiltonian MCMC, and Riemann Manifold Langevin and Hamiltonian Monte Carlo, \cite{neal2011mcmc,hoffman2014no,girolami2011riemann,CHANDRA2019315}. In such situations, canonical MCMC with random-walk proposals is an option, but their efficiency is  limited in the presence of  multiple modes.} 
  
 \textcolor{black} {Parallel tempering  MCMC provides efficient exploration of  multimodal  posteriors    \cite{marinari1992,geyer1995annealing} by featuring an ensemble of replica chains that are typically executed in parallel. The replica ensemble uses a temperature ladder to diffuse the likelihood, where the replicas with higher temperature values are more likely to accept weaker proposals,  enabling them from escaping local minima. Apart from this, we change the configuration in neighbouring replicas during sampling which further helps in providing a balance of exploration and exploitation. Each chain, $k$, has a stationary distribution of the form $\pi_k=p(\theta|\mathcal{D})^{1/\beta_k}$, where $\beta_k\in[1,\infty)$ is referred to as the replica's temperature. The temperature ladder $\boldsymbol\beta=(\beta_1,\ldots,\beta_K)$, is a $K\times 1$ vector of temperature values, where $K$ is the number of replicas. The elements of the temperature ladder have the property $\beta_1=1<\beta_2,\ldots,<\beta_K$, so that $k=1$ corresponds to the posterior. For values $k>1$, the stationary distribution  diffuses progressively, thus allowing the chain to escape from local modes.} 
 
 We propose the values of $\theta$ in each replica to swap with the values in neighbouring replicas so that the $\theta$'s from replicas $k>1$ have the potential to reach the replica corresponding to $k=1$,  for which the stationary distribution is the posterior.  In this way, parallel tempering explores the posterior via both local and global moves enabling them to explore multimodal distributions efficiently \citep{geyer1995annealing}.   Parallel tempering has been popular in geophysical inversion problems that have   complex 
 multimodal posterior distributions    \cite{sambridge2013parallel,sen1996bayesian,maraschini2010monte,sen2013global}.  More recently, we used parallel tempering   for landscape evolution models \cite{chandra2018PT-Bayes},   and three-dimensional joint inversion for mineral exploration \cite{Scalzo2019} which motivate their use in the proposed Bayesreef framework.

\section{Methodology: Bayesreef}

While deterministic models of long-term interactions of organisms in a marine ecosystem exist \cite{clavera2017process}, and software tools for coral reef evolution are available   \cite{Salles2018, barrett2017}, there are no probabilistic methods which combine these deterministic models with observed data. We note that such models have several or many non-unique solutions; however, even if we manage to constrain them to be unique, there remains uncertainty with the solution. \textcolor{black}{For example, we could estimate the Malthusian parameter in {\it pyReef-Core} as a   value of  0.5, which defines the population growth of the coralgal assemblages over a timeframe. We need to make a probabilistic statement which expresses the uncertainty of this estimate, such as the statement that the parameter lies between 0.25 and 0.87 with probability 0.90. To make such statements, we need a   consistent framework  which  fully accounts for different sources of uncertainty \cite{Jens2007UncertainEnvi}.} In addition,  there are several other factors of variability, such as limited, sparse, or missing geological data  \cite{hassan2009using,raje2012bayesian,Jens2007UncertainEnvi}.
 
In this section, we present the   \textit{Bayesreef} framework and demonstrate how heterogeneous sources of information can be combined to estimate the unknown parameters which govern coralgal vertical accumulation over time.  We assess the performance of our method by creating synthetic reef core ground-truth and compare with the \textit{pyReef-Core} output for coralgal vertical accumulation (prediction). We first discuss the creation of the synthetic core and then present the  \textit{Bayesreef} framework.

\begin{figure}[htb!]
\includegraphics[width=90mm]{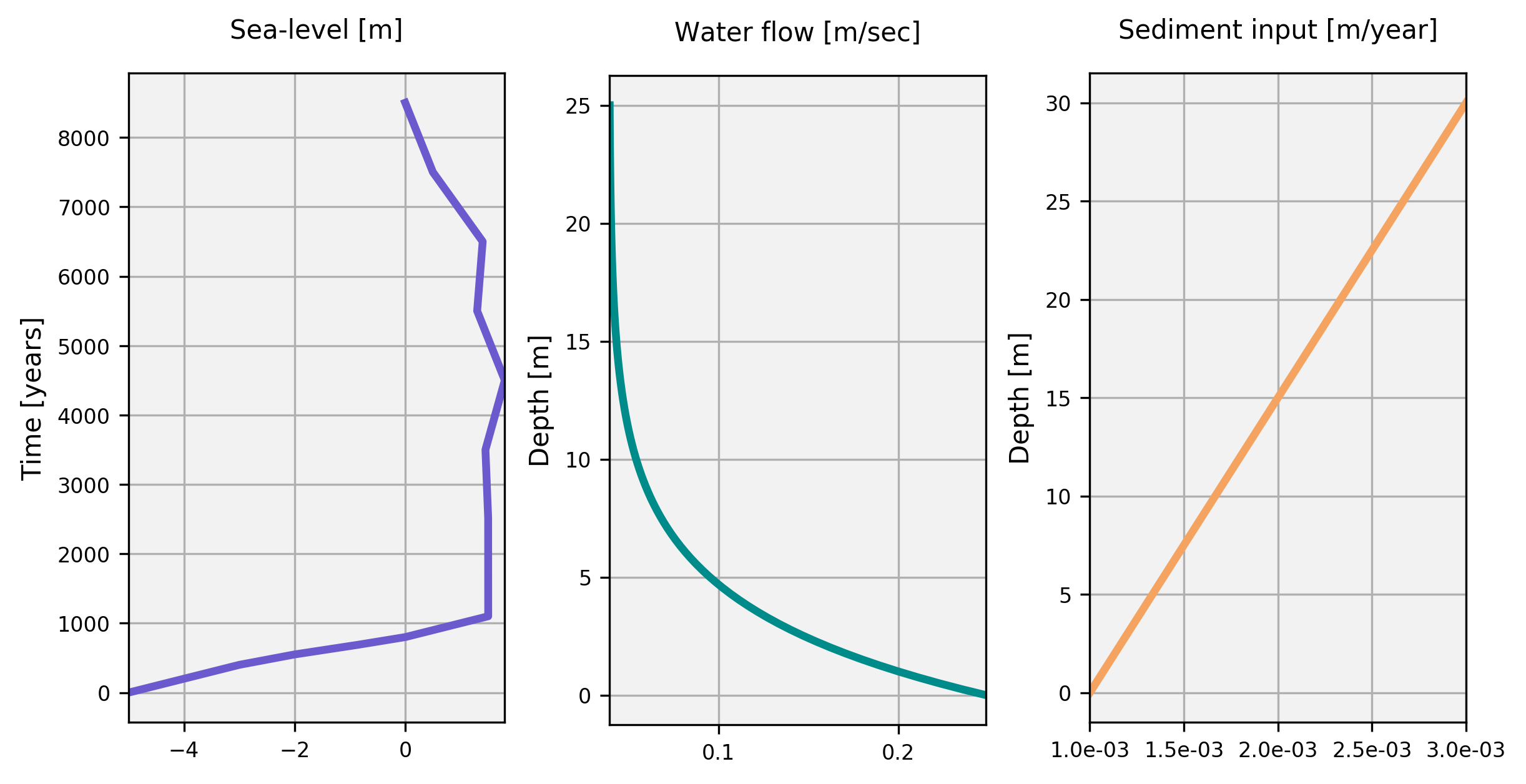}
\caption{Graph of the boundary conditions established for sea level (left), water flow velocity (centre), and sediment-input (right), used to create the synthetic ground-truth (Figure \ref{fig:workflow}) and  subsequent \textit{Bayesreef} experiments.} 
\label{fig:boundary}
\end{figure}

\begin{figure}[htb!]
\includegraphics[width=90mm]{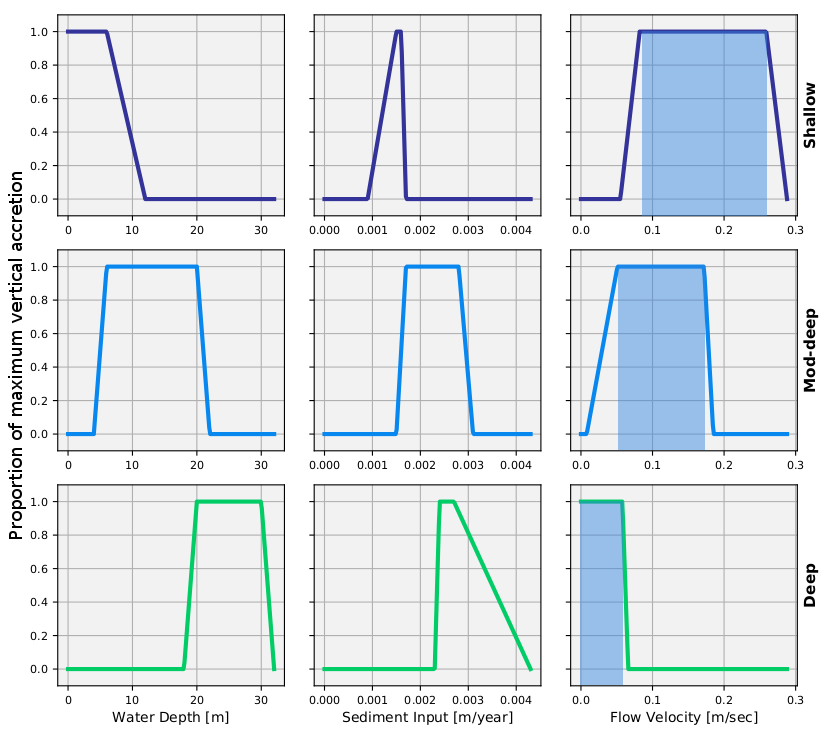}
\caption{Graph of the environmental threshold function for the shallow (upper), moderate-deep (centre), and deep (lower) assemblages characteristic of an exposed reef margin, interpreted from Dechnik\cite{Dechnik16}. The x-axis indicates the limitation on maximum vertical accretion for conditions outside the optimal, 100\% maximum growth window, indicated for clarity (blue translucent boxes) for the hydrodynamic energy threshold for all assemblages.  We present the maximum vertical accretion rate  for the respective assemblages  in Table \ref{tab:free-fixed-parameters},  used to create the synthetic ground-truth in Figure \ref{fig:workflow}.}
\label{fig:thresholds}
\end{figure}

\begin{table}[htb]
\begin{center}
\small 
\begin{tabular}{|l|p{7mm}p{7mm}p{7mm}p{7mm}|}
\hline
\textbf{Free parameters} & \multicolumn{4}{|c|}{\textbf{True values}}\\
\hline
\textbf{Population dynamics}  & \multicolumn{4}{|c|}{}\\
Malthusian parameter ($\varepsilon$) & \multicolumn{4}{|c|}{0.08}\\
AIM super-/sub-diagonals ($\alpha_s$) & \multicolumn{4}{|c|}{-0.03} \\
\hline
\textbf{Hydrodynamic energy} (m/s) & $f_{flow}^1$ & $f_{flow}^2$ & $f_{flow}^3$ & $f_{flow}^4$\\
Moderate-deep assemblage & 0.008 & 0.051 & 0.172 & 0.185\\

\hline \hline
{\bf Fixed parameters} & \multicolumn{4}{|c|}{\bf True values}\\ 
\hline
\textbf{Population dynamics} & \multicolumn{4}{|c|}{}  \\
AIM main diagonal ($\alpha_m$) & \multicolumn{4}{|c|}{-0.01} \\
\hline
\textbf{Hydrodynamic energy} (m/s) & $f_{flow}^1$ & $f_{flow}^2$ & $f_{flow}^3$ & $f_{flow}^4$\\
Shallow assemblage & 0.055 & 0.082 & 0.259 & 0.288\\
Deep assemblage & 0.000 & 0.000 & 0.058 & 0.066\\ 
\hline
\textbf{Sediment-input} (m/yr) & $f_{sed}^1$ & $f_{sed}^2$ & $f_{sed}^3$ & $f_{sed}^4$\\
Shallow assemblage & 0.0009 & 0.0015  &0.0016  &0.0017 \\
Moderate-deep assemblage & 0.0015 & 0.0017 & 0.0028 & 0.0031 \\
Deep assemblage & 0.0023 & 0.0024 & 0.0027 & 0.0043\\
\hline
Stratigraphic layer interval (yr) & \multicolumn{4}{|c|}{50} \\
\hline
\textbf{Maximum VA rates} (m/ka) & \multicolumn{4}{|c|}{} \\
Shallow assemblage &  \multicolumn{4}{|c|}{11} \\
Moderate-deep assemblage & \multicolumn{4}{|c|}{12} \\
Deep assemblage &  \multicolumn{4}{|c|}{9}\\
\hline
\end{tabular}
\end{center}
\caption{A summary of unknown (free) and fixed parameters used in \textit{Bayesreef} experiments. The true values of free parameters are used to obtain the synthetic ground-truth (Figure \ref{fig:synthcore}). Note that the true values of the hydrodynamic energy and sediment input exposure thresholds for each assemblage are graphically represented in Figure \ref{fig:thresholds}.  }
\label{tab:free-fixed-parameters}
\end{table}

\subsection{Creation of synthetic ground truth}
\label{sec_core_creation}

We create a  synthetic drill-core using  \textit{pyReef-Core}  that represents an idealised shallowing-upward fossil reef sequence with a catch-up growth strategy consistent with the Holocene evolution of several reefs globally,  shown in Figure~\ref{fig:synthcore}. If a coralgal assemblage is present, the type of coralgal assemblage is recorded (shown in green, light blue,  and dark blue); otherwise \textit{pyReef-Core} produces sediment deposition (shown in color \textit{orche}) in Figure \ref{fig:synthcore}. 
 
We represent the  information from the drill-core  in two forms, as  the depth-structure and the time-structure.  The depth structure records the coralgal assemblage present at various depths in the drill core, while  the time structure  records the time at which the  coralgal assemblage in the core was formed and both representations are present it {\it pyReef-Core}.
  
 The timeframe used to create the synthetic drill core was from 8.5 thousand years ago (ka), where water depth was likely to be 5~metres below sea level (MBSL), to present-day \cite{sloss2007holocene}. This period is within the take-off envelope for Holocene growth for outer-platform reefs, which ranges from 8.6 to 6.6~ka \cite{Hopleyetal07}. The initial reef surface was set to be 30~MBSL, which is consistent with the base of shallowing-upward sequences exhibited in drill cores from the Great Barrier Reef (GBR) and Indo-Pacific Reef \cite{salas2018holocene,Montaggioni05}. 

In the case of the synthetic reef-core, there are three types of coralgal assemblages produced in the \textit{pyReef-Core} simulation; shallow, moderate deep, and deep. They are consistent with those found on southern GBR reefs and capture the full extent of the shallowing-upward sequence in a high-energy and exposed setting. We set the initial populations for all three assemblages to zero. The initial conditions for the population dynamics, the hydrodynamic energy, sediment input and    maximum vertical accretion (VA) rates, for both fixed and free parameters, appear in Table~\ref{tab:free-fixed-parameters}.  We chose these values to mimic the thicknesses of facies and the timing of deposition when compared to a review of all Holocene drill cores \cite{Montaggioni05}. The VA rates for the three assemblages are defined based on a full analysis of all Indo-Pacific reef drill cores \cite{Montaggioni05, Dechnik16}.

We assume that flow velocity and related hydrodynamic energy is an exponentially decreasing function of depth (Figure~\ref{fig:boundary}).  The flow velocity varies from extremely low, laminar flow ($<$4~cm/sec) on the deep forereef ($>$30~m depth) to mean flow speeds of 20-30~cm/sec in $<$1~m depth \cite{sebens2003effects}. This relationship has been validated by lab and field studies \cite{comeau2014water,fulton2005wave, davies1983growth} and based on them, we restrict the maximum flow velocity in   {\it pyReef-Core} to 30~centimeters/second.

The environmental threshold functions for the ground-truth depicted in Figure \ref{fig:thresholds} entail the growth response of coralgal assemblages to changing depth, sediment input, and hydrodynamic energy.  We construct the  \textit{pyReef-Core} model so that maximum VA rates for the respective assemblage only reach under optimal conditions. Elsewhere, growth is proportional to the environmental factor determined by exposure threshold functions \cite{Salles2018} (Figure \ref{fig:thresholds}). 
 
The depth exposure thresholds for each coralgal assemblage are well-defined in the literature \cite{Dechnik16,dechnik2017evolution}; however, there is little to no data on the optimal growth environment in relation to other environmental factors which is not at the species level and certainly not on greater-than-decadal timescales. Therefore, we manually estimate the threshold functions for sediment input and hydrodynamic energy for the creation of the synthetic ground-truth (Figure~\ref{fig:thresholds}). This manual estimation is done mainly by trial and error, until we obtain the desired shallowing-upward sequence that accurately reflects the expected shift from deep to moderately-deep assemblages at $\sim$15$-$20~MBSL, and from moderately-deep to shallow assemblages at $\sim$6~MBSL \cite{Cabiochetal99,Dechnik16} (Figure~\ref{fig:synthcore}). 

The simulations in {\it pyReef-Core} use a depth-dependent sediment input function to approximate the spatial variation in sedimentation rate resulting from hydrodynamic conditions. Following the same approach as for the definition of the hydrodynamic energy, we use a sedimentation-depth relationship conceptualised by Chappell \cite{chappell1980coral} to simulate sediment deposition in this study (Figure \ref{fig:boundary}).

We simplified the vertical accommodation, which refers to the vertical space available for potential reef accumulation,   to be a function of Holocene sea-level changes and vertical coral reef growth only. The Holocene relative sea-level (RSL) curve for the Australian East Coast is used as the sea-level boundary condition, presented in Figure \ref{fig:boundary} \cite{sloss2007holocene}. The data indicates a RSL history characterised by a mid-Holocene highstand of 1.8~m at $\approx 4$ka,  before returning slowly to present sea-level. We obtain the temporal and vertical evolution of the produced synthetic core   at user-defined intervals and depend on each assemblage production rate that varies based on the aforementioned input parameters.

\begin{figure*}[htb!] 
\includegraphics[width=180mm]{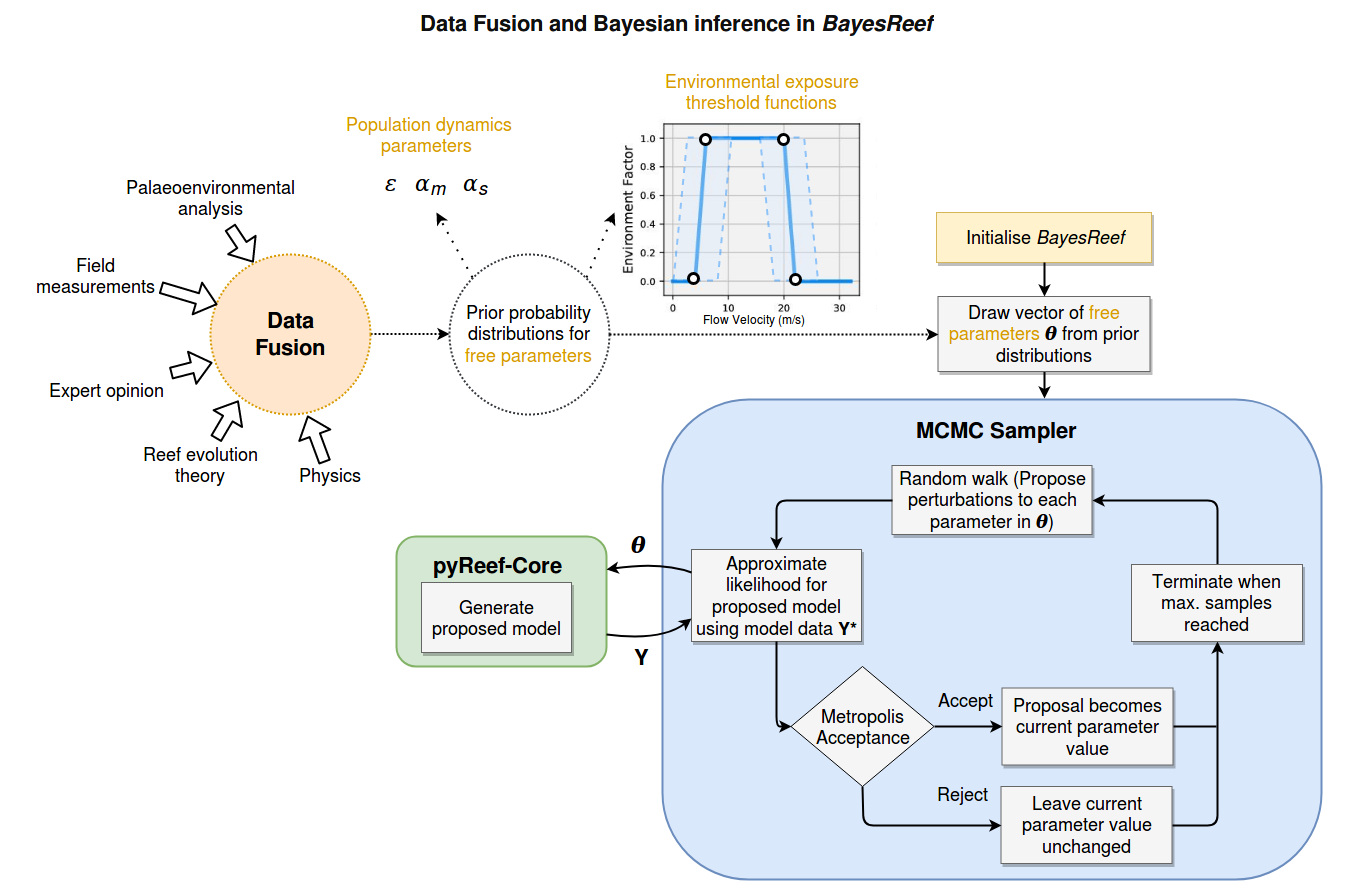}
\centering
\caption{ \textit{Bayesreef} framework with MCMC sampling for inference of free parameters in the \textit{pyReef-Core} model. The prior probability distribution represents the subjective beliefs about free parameters before taking into account the evidence or data.   The priors are used to propose the initial vector of free parameters (for population dynamics and environmental exposure threshold functions) to be used in the MCMC sampler.   
} 
\label{fig:Bayesreef}
\end{figure*}   

\subsection{Data and parameters}
\label{sec_likl}

To demonstrate the convergence of MCMC sampling for Bayesreef, we consider the synthetic ground-truth reef-core as observed data.  In this way, we conduct experiments to find if we can recover the true parameter values used to create the synthetic ground-truth.  The observed coralgal assemblages given in the reef-core is given as   $\mb y_{\mathbb D}^{o}=(y_{\mathbb D,1}^o,\ldots,y_{\mathbb D,D}^o)$, where $y^o_{\mathbb D,d}$ is the type of coralgal assemblage at depth $d$, for $d=1,\ldots,D$. 
The superscript $o$ denotes  observed data, while the subscript ${\mathbb D}$  denotes that the data has a depth structure.

We denote the coralgal assemblage  data as $\mb y_{\mathbb T}^{o}=(y_{\mathbb T,1}^o,\ldots,y_{\mathbb T,T}^o)$, where the subscript ${\mathbb T}$ is to denote that the data has a time structure, and  $y^o_{\mathbb T,t}$ is the type of coralgal assemblage observed at time $t$, for $t=1,\ldots,T$. We note that there can be many time structure combinations which give rise to the same depth structure, but not vice-versa.  Thus, the time structure contains more information than the depth structure.  Hence there is only a one-to-one correspondence between the two if the rate of coralgal growth is constant over the entire timeframe in question. We show how using this knowledge of the time structure allows us to constrain the parameter space. For conciseness, we drop the subscripts ${\mathbb D}$ and ${\mathbb T}$ and develop the general model for the data.

Given the data $\mb y^{o}$, inference about the unknown parameters $\bs\theta$, proceeds via the posterior distribution, $p(\bs\theta|\mb y^o)$.  The denominator in Equation~(\ref{eqn_Bayes}) is merely a normalizing constant and does not depend upon $\bs \theta$, so that Equation~(\ref{eqn_Bayes}) is often expressed as 
\[
p(\bs\theta|\mb y^o) \propto p(\mb y ^{o}|\bs\theta)\times p(\bs\theta).
\]
where $p(\cdot)$ represents a probability density function, $p(\mb y^{o}|\bs\theta)$ is the likelihood function, and $ p(\bs\theta)$ is the prior distribution. 

The unknown parameters of interest ($\bs\theta$) are; the Malthusian parameter ($\epsilon$); the elements of the assemblage interaction matrix, denoted by the matrix $\bs A$; four critical points that define the hydrodynamic energy exposure threshold function for each assemblage,  ${\bf f}_{i,flow}=(f^1_{i,flow}, ..., f^4_{i,flow})
$where $i$ denotes the assemblage type; and four critical points defining the exposure threshold function of sediment input
${\bf~f}_{sed}=(f^1_{sed},~..., f^4_{sed}).
$.  Hence, the vector of parameters for  inference is $\bs\theta=(\epsilon,\mb A, \mb f_{1,flow},\mb f_{2,flow},\mb f_{3,flow}, \mb f_{sed})$.  The details about  the ``true" value of these parameters used in producing the  synthetic ground-truth, $\mb y^o$ appear in Table \ref{tab:free-fixed-parameters}.\\ 

\subsection{Likelihood}

The likelihood is a probabilistic model  of the data-generating process given some parameters $\boldsymbol\theta$. It is used to describe how likely the observed data are if we knew the true value of $\bs\theta$.
The existence and type of  coralgal assemblage that grows at various points in time are random variables that take on a discrete number of outcomes; either coral grows or it does not (in which case we observe sediment). For instance, if the coral grows, then the assemblage type is either shallow, moderate-deep or deep; and hence, the likelihood function should reflect the discrete nature of these outcomes.

The observations in Bayesreef are discrete; namely the type of coralgal assemblage that forms at a given point in time and/or depth. Therefore we choose the product of  independent multinomial distributions for the likelihood function. There is no closed form expression for $p(\mb y^o|\bs \theta)$, however there is a deterministic relationship between $\bs\theta$ and the proportion of  various assemblages at equally spaced points in time. We  denote these proportions  by
$\bs\Pi=(\bs\pi_1,\ldots,\bs\pi_{T})$, where $\bs\pi^m_t=(\pi^m_{t1},\ldots,\pi^m_{tK})$, for $t=1,\ldots,T$,  with $\sum_{k=1}^{K}\pi_{tk}=1$ given coralgal  assemblage $m$.  These proportions are time-varying and depend deterministically on $\bs\theta$ via the {\it pyReef-Core} forward model. 

This deterministic relationship between $\bs\theta$ and $\bs \Pi$, allows us to evaluate the likelihood because  $p(\mb y^o|\bs \theta)=p(\mb y^o|\bs \Pi)$.
We consider  the elements of the $T\times K$ matrix $\bs\Pi$, namely $\pi_{tk}$, to be the probability that assemblage $k$ is present at time $t$, and use these probabilities as inputs for the multinomial likelihood. So that 
\begin{equation}
\Pr(\mb y^{o}| \bs\theta^m)=\prod_{t=1}^{T}\prod_{k=1}^{K} \pi^{z_{tk}}
\label{eq_multi_like}
\end{equation}
where $z_{tk}=1$ if $y_t=k$, and $z_{tk}=0$, otherwise.


\subsubsection{Priors for $\varepsilon$ and $A$}

We made some modifications in  \textit{pyReef-Core} to simplify it by reducing the number of unknown parameters to make the Bayesreef framework computationally feasible.  First, we assume that the value of the Malthusian parameter is equal for all coralgal assemblages and  we place an uninformative prior on $\varepsilon$, so that $p(\varepsilon)\sim~U[0,0.15]$. Second, we assume the assemblage interaction matrix ($\mb A$),  to be symmetric and block diagonal with equal diagonal elements, so that
\[
\mb A
=
\begin{bmatrix}
  \alpha_{m} & \alpha_{s} & 0\\
  \alpha_{s} & \alpha_{m} & \alpha_{s} \\
  0 & \alpha_{s} & \alpha_{m}
\end{bmatrix}
\]

The zeros in the matrix indicate conditional independence between two assemblages that are not close together in space   and time \cite{eppley1972temperature}. The prior for $p(\alpha_{m})$, and $p(\alpha_{s})$ is $U[-0.15,0]$, which represent the interaction between the different coralgal assemblages.

There are additional restrictions imposed upon combinations of $\varepsilon$ and $A$. We use the Runge-Kutta-Fehlberg  (RKF-45) method as the numerical ordinary differential equation (ODE) solver; however, it becomes linearly unstable when the magnitude of the difference between $\varepsilon$ and $\alpha_{m}$ or $\alpha_{s}$ is too large. This is because RKF-45 uses an adaptive step-size, which has limited stability when dealing with stiff equations \cite{butcher1996history}. We address this issue and ensure stability, by limiting the range of $\varepsilon$, $\alpha_{s}$ and $\alpha_{m}$  to the interval (0.15 and -0.15), respectively (Table \ref{tab:priors}).

\subsubsection{Priors for hydrodynamic energy and sediment input}
\label{sef_flow_priors}
The parameters that define the sediment input exposure threshold function $\mb f_{sed}$, and the hydrodynamic energy exposure threshold function  $\mb f_{flow}$, serve as constraints that restrict vertical growth to only occur within a range of values for these environmental stressors.
There are four parameters for each assemblage that define the exposure threshold function to sediment-input, and four parameters that define the exposure threshold function to the hydrodynamic energy ( flow velocity). Hence, if there are three assemblages,  then the total number of parameters which define the sediment-input and flow velocity is 24 (12 parameters for $\mb f_{sed}$ and another   12 parameters for $\mb f_{flow}$).  In the case of six assemblages, there are 48 parameters  (24 parameters for $\mb f_{sed}$ and another   24 parameters for $\mb f_{flow}$). In both cases, 3 other parameters are always included, i.e. the Malthusian and two AIM parameters.  

Based on the sediment-depth and flow-depth relationships established in Section \ref{sec_core_creation}, the maximum flow velocity is 0.3~meters/second (m/s) and the maximum sediment-input is 0.005~meters/thousnad years (m/kyr). The limits on these environmental factors have been informed by physics and prior knowledge of reef systems. The prior distributions for all elements of $\mb f_{i,flow}$ and  $\mb f_{sed}$ ensure an ordering.  For example, $\mb f_{i,flow}=(f^1_{i,flow_1}, f^2_{i,flow},f^3_{i,flow},f^4_{i,flow})$ has a  prior distribution such that

$$P(f_{i,flow})=P(f^1_{i,flow})
\times
\prod_{j=2}^4 p(f^j_{i,flow}|f^{j-1}_{i,flow})$$
 with $f^j_{i,flow}\sim U(f^{j-1}_{i,flow},0.3]$.

\begin{table}[htb]
\begin{center}
\small 
\caption{Prior distributions and the proposal standard deviation ($\sigma$) for the Metropolis-Hasting kernel in the MCMC sampling  scheme in \textit{Bayesreef}. Note that proposal standard deviation is 1\% of the width of the absolute range of the prior.} 

\label{tab:priors}
\begin{tabular}{ c c c }
\hline
 Parameters  &  Priors &   Step-size ($\bs\sigma$) \\
\hline
$\mb f_{flow}$ & $P(f_{flow_1})
\times
\prod_{j=2}^4p(f_{flow_j}|f_{flow_{j-1}})$ & 0.00300 \\
$\mb f_{sed}$ & $P(f_{sed_1}) 
\times
\prod_{j=2}^4p(f_{sed_j}|f_{sed_{j-1}})$ & 0.00005\\
$\varepsilon$ & U$[ 0.00 ,0.15 ]$ & 0.00150 \\
$\alpha_m$ & U$[ -0.15, 0.00 ]$ & 0.00150 \\
$\alpha_s$ & U$[ -0.15, 0.00 ]$ & 0.00150 \\
\hline
\end{tabular}
\end{center}
\end{table}
There is little to no quantitative data collected on the growth response of coral species (let alone coralgal assemblages) in response to different values of flow velocity and sediment-input. As such, we cannot incorporate more information into the prior. The relaxed constraint we chose gives the best chance for coralgal vertical accretion to be simulated in the model by avoiding situations where a threshold is too narrow to allow growth to occur at all. Therefore, they formally express the state of limited knowledge regarding long-term effect of hydrodynamic energy and sediment flux regimes on coral growth. The details regarding their corresponding prior distributions appear in Table \ref{tab:priors}.

\subsection{Estimation via Bayesian inference} 
\subsubsection{Proposal constraints}

 The elements of $\bs\theta$, which correspond the flow velocity and sediment-input parameters, need to be ordered, as described in Section~\ref{sef_flow_priors} and shown in Figure \ref{fig:thresholds}.   To impose these constraints during  MCMC sampling, we use random-walk proposal distribution, but then we sort selected parameters to preserve order in the flow velocity and sediment-input parameters.  We can view this as a bi-level constraint parameter optimisation problem where the first level constraints are the parameter limits, and second-level constraint is the sorting of certain adjacent parameters.  Hence, after we check the limits for each proposed value, we sort the adjacent parameters, which is a simple but effective method while efficiently exploring the posterior \citep{jasra2005}. Algorithm \label{alg:cons} describes the first-level constraints for the three GLV (Malthusian and community interaction) parameters and heuristic for bi-level constraints given for the flow and sediment-input parameters for the respective assemblages.  Creating proposal distributions for generating such constraints adds further challenges for  MCMC sampling.

 \begin{algorithm}[htp]
 \SetAlgoLined
 \small
 \KwResult{Valid proposal vector}
 Get current state vectors:\\
 i. GLV parameters(\textbf{v})\\
 ii. Flow parameters for all assemblages  (\textbf{x})\\
 ii. Sediment-input parameters for all assemblages (\textbf{w}) \\
 Get minimum and maximum limit for all parameters. Note that flow and sediment-input parameters used the same limits (uniform priors) for all assemblages as shown in Table \ref{tab:priors}.

  First three parameters \\
 
  \textbf{GLV parameters} \\
   1. First level constraint: check limits  \\
   \For{i in each GLV}{
   v-new[i] = \textbf{v}[i] + \textit{Gaussian-Noise()} \\
   \eIf{\textbf{v}[i] $\geq$ v-min[i] and \textbf{v}[i] $<$ v-max[i] }{
   \textbf{v}[i] = v-new[i]   \\
   }{
   Set \textbf{v}[i] to previous value \\
   }
  }
 
  \textbf{Flow parameters} \\
  \For{i in each Assemblage}{ 
   1. First level constraint: check limits   \\
   \For{j in each Flow}{
   x-new[i][j] = x[i][j] + \textit{Gaussian-Noise()} \\
   \eIf{x[i][j] $\geq$ x-min[j] and x[i][j] $<$ x-max[j] }{
  x[i][j] = x-new[i][j]   \\
   }{
   Set x[i][j] to previous value \\
   }
  }
 
  2. Second level constraint:\\
  Sort  to ensure:  x[i][0] $<$x[i][1] $<$ x[i][2] $<$ x[i][3] \\
  }
 
  \textbf{Sediment-input parameters} \\
   \For{i in each Assemblage i}{ 
  3. First level constraint: check limits  \\
   \For{j in each Sediment-input}{
   w-new[i][j] = w[i][j] + \textit{Gaussian-Noise()} \\
   \eIf{w[i][j] $\geq$ w-min[j] and w[i][j] $<$ w-max[j] }{
     w[i][j] = w-new[i][j]   \\
   }{
   Set w[i][j] to previous value \\
   }
  }
 
 4. Second level constraint:\\
  Sort  to ensure:  w[i][0] $<$w[i][1] $<$ w[i][2] $<$ w[i][3] \\

  }

  \caption{Heuristic for implementing bi-level constraints in proposals  }
  \label{alg:cons}
 \end{algorithm}

\subsubsection{MCMC Sampling} 
\label{bayes_inference}
As noted earlier, we take a Bayesian approach and use the posterior distribution $p(\bs\theta|\mb y^o)$ and make inference regarding $\bs\theta$. We denote the predictive distributions of an assemblage at time $T+1$ to be  $\Pr(y^*_{T+1}|\mb y^o)$; where the notation $y^*$ indicates an unobserved prediction and $\mb y^o$ indicates the observed value from time $t=0$ to time $t=T$. We integrate over all possible values of $\bs\theta$, 
\begin{equation}
\Pr(y^*_{T+1}|\mb y^o)=\int \Pr(\mb y^*|\mb y^o,\bs\theta)p(\bs\theta|\mb y^o)d\bs\theta\\
\label{eqn_int_mcmc}
\end{equation}

Then, the integral in Equation \eqref{eqn_int_mcmc} is approximated by 
\begin{equation}
\Pr(y^*_{T+1}|\mb y^o)
\approx\frac{1}{M}\sum_{j=1}^M \Pr (y_{T+1}^*|\mb y^o,\bs\theta^{[j]})
\end{equation}
where $\bs\theta^{[j]}$ is  drawn from the posterior distribution $p(\bs\theta|\mb y^o)$.  We use MCMC sampling to obtain these draws, where the transition kernel is a random-walk (RW) Metropolis-Hastings kernel with a proposal distribution $q(.)$, which is $\phi(\bs \theta^c,\Sigma)$; where $\bs\theta^c$ is the current value of $\bs \theta$ in the chain and $\phi(\bs\mu,\Sigma)$ is the multivariate normal probability density function with mean vector $\bs\mu$ and covariance matrix $\Sigma$.

Algorithm \ref{alg:alg1} presents the single-chain MCMC and Algorithm \ref{alg:algpt} presents parallel tempering MCMC. We note that both algorithms  compute the   acceptance probability for within chain proposals by  \begin{equation}
\vcenter{\begin{align}p_{accept} =
    \textrm{min}\left\{ 1,\dfrac{p\left(\mb y|\bs\Pi^{[p]}\right) p\left(\bs\theta^{[p]}\right)}
    {p\left(\mb y|\bs\Pi^{[i-1]}\right) p\left(\bs\theta^{[i-1]}\right)}
     \cdot \dfrac{q\left(\bs\theta^{[i-1]}|\bs\theta^{[p]}\right)}
     {q\left(\bs\theta^{[p]}|\bs\theta^{[i-1]}\right)} \right\} \end{align}
     \nonumber \\
 }\label{eqn_accept}
 \end{equation}
 where, $p(\mb y|\bs\Pi)$ is given by Equation~\ref{eq_multi_like}. Note that if any of the constraints for $\varepsilon$ and the elements of $A$ are not met, the acceptance probability is set to zero.   The  Bayesreef framework  is shown in Figure \ref{fig:Bayesreef} which highlights  single-chain MCMC sampling shown in Algorithm 2 where Algorithm 3 can also be used.

\begin{algorithm}[htp]
\SetAlgoLined
\small
\KwResult{Posterior distribution}
 Initialise chain, $\bs\theta=\bs\theta^{[0]}$: \\
 \indent Draw $\bs\theta^{[0]}$ from the joint prior distribution $\bs\theta^{[0]}\sim p(\bs \theta)$

  \For{i until  $\mbox{Iter}_{\max}$ }{
  
   1. Propose  $\bs\theta^{[p]}|\bs\theta^{i-1}\sim q(\bs\theta^{[i-1]})$ where $q(.)$ is the selected proposal distribution (Algorithm 1). \\ 
  2.  Given  $\bs\theta^{[p]}$, use \textit{pyReef-Core}  to predict the  reef-core in order to evaluate the likelihood (Equation \ref{eq_multi_like}) \\
  
 3. Calculate acceptance probability (Equation \ref{eqn_accept}) \\
  
 Draw $\alpha$ from a uniform distribution,  $\alpha \sim U(0,1)$  \\
  \eIf{ $p_{accept} < \alpha$}{
  set current value as proposed value, $ \bs\theta^{[i]} =\bs\theta^{[p]}$ \\
   }{
    current value remains unchanged, $\bs\theta^{[i]} =\bs\theta^{[i-1]} $ \\
  }
 }

 \caption{Single-chain MCMC sampling  }
 \label{alg:alg1}
\end{algorithm}

In Algorithm  \ref{alg:alg1}, single-chain MCMC   draws the initial   values   $\theta_m$  from the prior $p(\theta)$ and moves onto the sampling stage.  We update the chain when the proposal is accepted/rejected using the  Metropolis-Hastings acceptance criterion given by Step 3, where we add the accepted proposal to the posterior distribution. We repeat the procedure until the termination condition is satisfied as given by the maximum number of iterations (samples), $\mbox{Iter}_{\max}$.
    
\textcolor{black}{  
In Algorithm \ref{alg:mhptmcmc},  parallel tempering MCMC  uses an ensemble of replica with a  temperature ladder to provide  balance between exploration and exploitation during sampling. The replicas with higher temperature values provide exploration while those with lower temperature values enforce exploitation. In the beginning, the number of replicas $M$, the maximum  number of iterations, $\mbox{Iter}_{\max}$,   and the temperature ladder $\bs\beta=(\beta_M,\ldots,\beta_1)$ is defined. We draw the initial values of $\theta_m$ for $m=1,\ldots,M$ from the prior $p(\theta)$, and move onto the sampling stage.  We update each replica $\theta$   when the respective proposal is accepted/rejected using the  Metropolis-Hastings acceptance criterion given by Step 1.3. Once all the replicas have sampled,  we check if we can swap the neighbouring replicas using Metropolis-Hastings criterion (Step 2.2).  We repeat the procedure until the termination condition is satisfied as given by the maximum number of iterations.  }
\begin{algorithm}[htp]
\small
\caption{\textcolor{black}{ Parallel tempering MCMC sampling}}\label{alg:mhptmcmc}

\KwResult{Posterior distribution}
\textcolor{black}{ i. Set maximum number of iterations  ($\mbox{Iter}_{\max}$),  the number of replicas ($M$), the percentage of iterations for parallel tempering phase, and the temperature ladder $\bs\beta=(\beta_M,\ldots,\beta_{1})$. \\
ii. Initialize replica $\theta_m=\theta_{m^{[0]}}^{[0]}$,  for $m=1,\ldots,M$.\\
iii.  Set the current values;   $\theta_m^c=\theta_{m^{[0]}}^{[0]}$, and   $m^c=m^{[0]}$\;  }
 
\For {$k=1,\ldots,$ $\mbox{Iter}_{\max}/M$}{ 
 
 \For {$m=1,\ldots,M$   }{ 
 \noindent
  
 Step 1: Replica sampling \\
 1.1  Propose  $\bs\theta^{[p]}|\bs\theta^{i-1}\sim q(\bs\theta^{[i-1]})$ where $q(.)$ is the selected proposal distribution (Algorithm 1)\\
 1.2   Given  $\bs\theta^{[p]}$, use \textit{pyReef-Core}  to predict the  reef-core in order to evaluate the likelihood (Equation \ref{eq_multi_like}) \\
1.3 Compute acceptance probability (Equation \ref{eqn_accept})\; 
 
Draw $u$ from uniform distribution, $u\sim U[0,1]$\\
\eIf{$u<\alpha$}{
set current value as proposed value, $\theta^{[k]}=\theta^p$\;
}{
 current value remains unchanged, $\theta^{[k]}=\theta^c$\;
}
 }
 \For {$m=1,\ldots,M-1$   }{
 Step 2: Replica Exchange \\
 
 2.1 Select a pair of neighbouring replica, $[\theta_m, \theta_{m+1}$] \\
  
2.2 Compute acceptance probability \\

Draw $u$ from uniform distribution, $u\sim U[0,1]$\\
 
\eIf{$u<\alpha$}{
exchange neighbouring replica,  $\theta_m \leftrightarrow \theta_{m+1}$
}

}
  }
   \label{alg:algpt}
 \end{algorithm}

\subsubsection{Adaptive proposal  for GLV parameters}

We investigate  the use of two different covariance matrices for the 3 GLV parameters. The first is to fix $\Sigma$ to be diagonal, so that $\Sigma=\mbox{diag}(s^2_1,\ldots, s^2_P)$; where $s_j$ is the step size of the $j^{th}$ element of the parameter vector $\bs\theta$ given in Table~\ref{tab:priors} for $j=1,\ldots,P$,  and $P$ is the length of $\bs\theta$.

The second covariance matrix allows for the dependency between elements of $\bs\theta$ and changes throughout the sampler and known as adaptive random-walk proposal distribution \citep{haario_adaptive_2001}.  Here, we adapt the elements of the step-size $\Sigma$ for the proposal distribution   using the sample covariance of the current chain history:  $\Sigma = \mbox{cov}(\{\bs\theta^{[0]}, \ldots, \bs \theta^{[i-1]}\}) + \mbox{diag}(\lambda_1^2,\ldots,\lambda_P^2)$; where $\bs\theta^{[i]}$ is the $i^{th}$ iterate of $\bs\theta$ in the chain and $\lambda_j$ is the minimum allowed step-size for each parameter $\theta_j$.  Although $\Sigma$  depends on the history of the chain, rather than just on the current values which make them not Markov, the proposals such as these with  \textit{vanishing adaptation} have guarantees of ergodicity in the long-term limit, allowing the adapted chain to sample from the desired distribution \citep{roberts_coupling_2007}. The rest of the parameters make the majority of the parameters in Bayesreef, which need second-level constraints and can only be created using Algorithm 1.

\section{Experimental design and results} 

We investigate the performance of \textit{Bayesreef} using the multinomial likelihood function outlined in the previous section for the time and depth structure and their associated time-based and depth-based likelihoods as the basis of prediction and evaluation of results. We use two different MCMC sampling methods for synthetic - reef-core and apply on a selected real-world application from the Great Barrier Reef. The experiments follow the plan below. 

\begin{enumerate}
 
    \item \textcolor{black}{Single-chain MCMC sampling (Algorithm 2) for two free parameters   for time structure versus depth structure setting. }
    \item \textcolor{black}{Single-chain MCMC sampling (Algorithm 2) for four free parameters for time structure versus depth structure setting. }
        \item \textcolor{black}{Single-chain MCMC sampling (Algorithm 2) for three free parameters  to evaluate adaptive random-walk and random-walk proposals for  depth structure setting. }
         \item \textcolor{black}{Convergence diagnosis for experiments above (Part 3). }
         \item \textcolor{black}{Compare single-chain MCMC  (Algorithm 2) with parallel tempering MCMC (Algorithm 3) for the 27 free parameters.}
         
         \item  \textcolor{black}{Use parallel tempering MCMC sampling (Algorithm 3) for application to  a fossil reef core from \textit{One Tree Reef}  from the southern Great Barrier Reef that features six assemblages; hence, 51 free parameters.}

\end{enumerate}

We note that all of the above steps use synthetic reef-core for experiments, except for the last step. The performance of \textit{Bayesreef} is measured in two ways: by its ability to recover the reef-core structure of coralgal assemblage composition and transitions when compared with the ground truth. In the case of the synthetic reef-core experiments, the evaluation is also by the estimation accuracy of the parameters used to create the reef-core. Note that the two performance measures are related but not the same. It could be that the same coralgal assemblage reef-core is predicted by different combinations of the parameters.


All experimental runs feature a 15\% burn-in period, where the samples (iterates) are not included in the posterior which is standard for MCMC sampling.  It takes about $\sim$11-12 hours to run 10,000 samples  using a a single core from  Intel Core i7-8700 Processor (6~Cores, 12~MB cache, 4.6~GHz).

\begin{table}
\begin{center}
\small
\begin{tabular}{|p{5mm}|p{7mm}|p{8.5mm}p{8.5mm}p{4mm}|p{8.5mm}p{8.5mm}p{4mm}|}
\hline
  Parameters  &  True value & \multicolumn{3}{c|}{  Time likelihood} & \multicolumn{3}{c|}{{\bf Depth likelihood}}\\
&& {\bf Mean  }& {\bf Mode } & {\bf AR (\%)} & {\bf Mean} & {\bf Mode} & {\bf AR (\%)}\\
\hline
\multicolumn{8}{|c|}{\textbf{2 free parameters}}\\
\hline
$\alpha_s$ & -0.03 & -0.029 & -0.035 & \multirow{2}{*}{36.7} & -0.033 & -0.043 & \multirow{2}{*}{61.4}\\
$\varepsilon$ & 0.08 & 0.081 &   0.082  &    & 0.136 & 0.133 &    \\
\hline
\multicolumn{8}{|c|}{\textbf{4 free parameters}}\\
\hline
$f_{flow}^1$ & 0.008 & 0.014 & 0.037 & \multirow{4}{*}{31.1} & 0.018 & 0.041 & \multirow{4}{*}{78.2} \\
$f_{flow}^2$ & 0.051 & 0.049 & 0.042 &    & 0.041 & 0.041 &     \\
$f_{flow}^3$ & 0.172 & 0.116 & 0.049 &    & 0.131 & 0.116 &    \\
$f_{flow}^4$ & 0.185 & 0.227 & 0.296 &    & 0.201 & 0.213 &     \\
\hline
\end{tabular}
\caption{Results showing posterior mean, mode and acceptance rate (AR) for parameter estimates from time and depth-based likelihood with two and four free parameters. 
}
\label{tab:summ-stats}
\end{center}
\end{table}
\begin{figure*}[htb!]
\centering
  \subfloat[]{\includegraphics[width=57mm,valign=t]{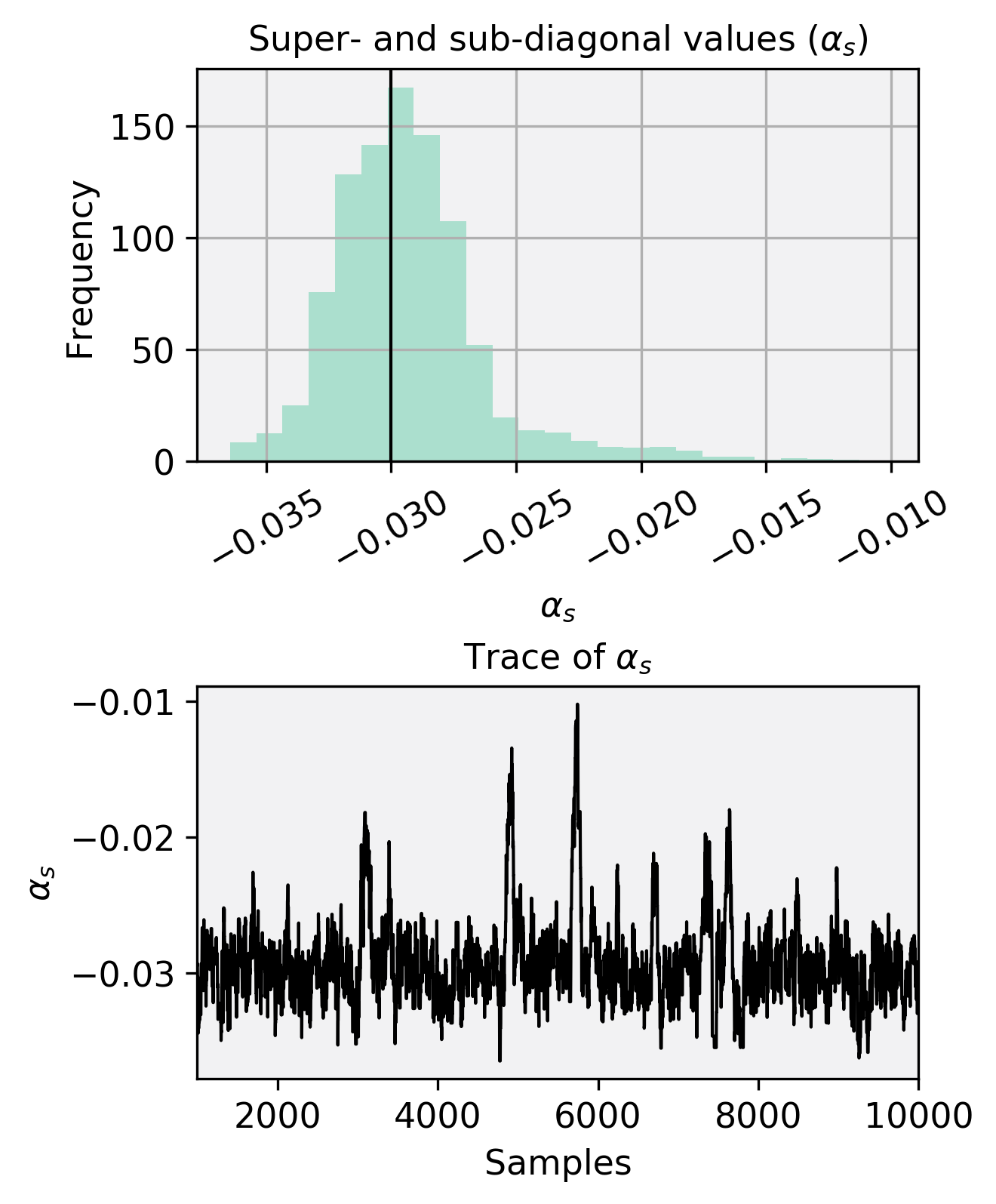}\label{fig:2p-t-ay}}
  \subfloat[]{\includegraphics[width=57mm,valign=t]{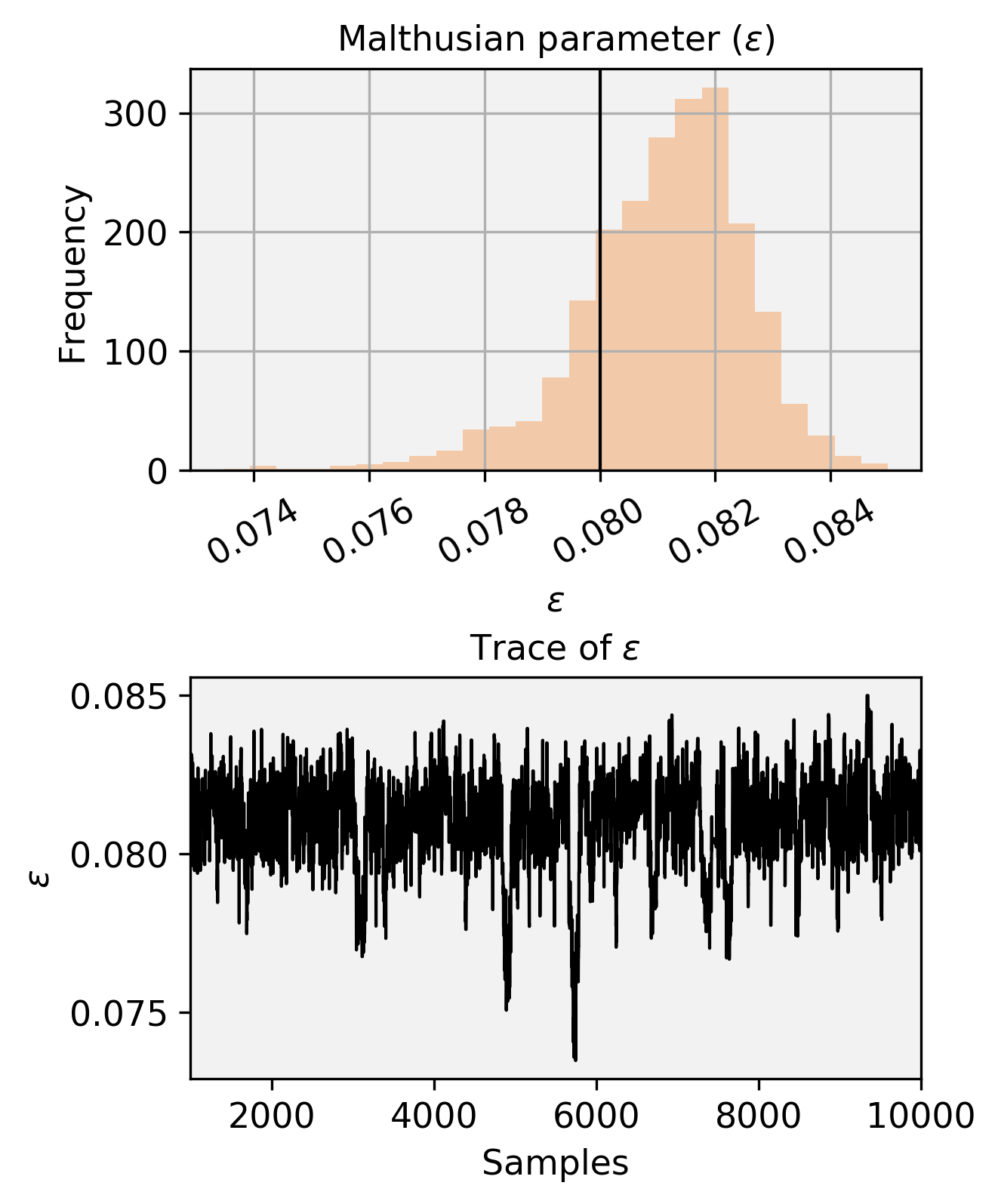}\label{fig:2p-t-malth}}
  \subfloat[]{\includegraphics[width=66mm,valign=t]{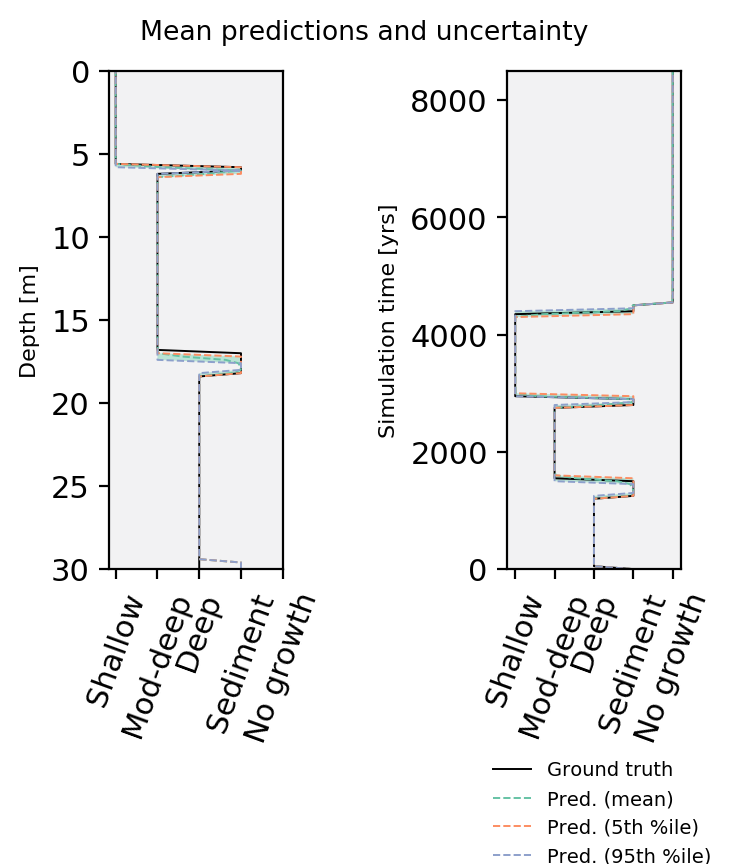}\label{fig:2p-t-mcmcres}}\\
  \subfloat[]{\includegraphics[width=57mm,valign=t]{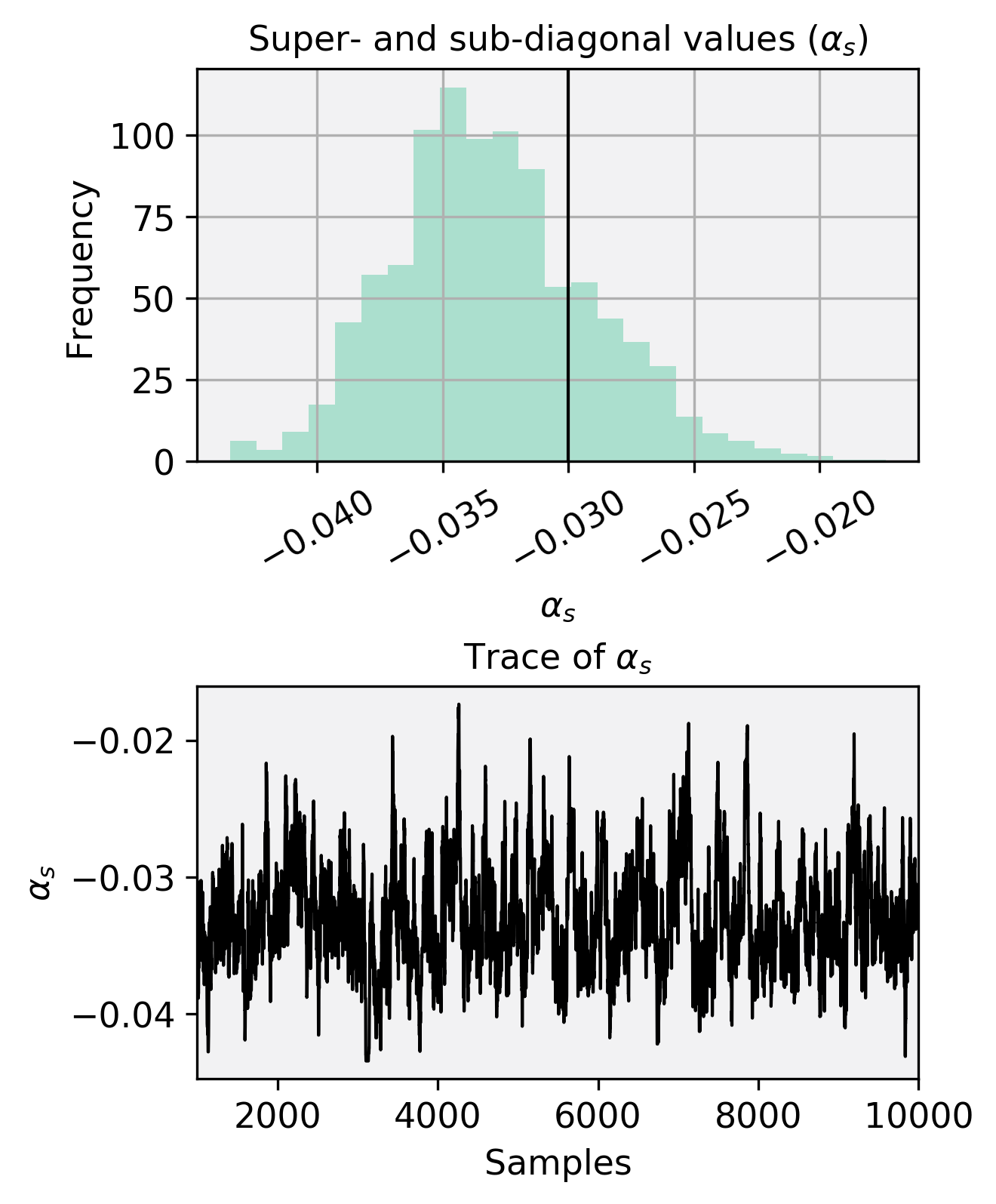}\label{fig:2p-d-ay}}
  \subfloat[]{\includegraphics[width=57mm,valign=t]{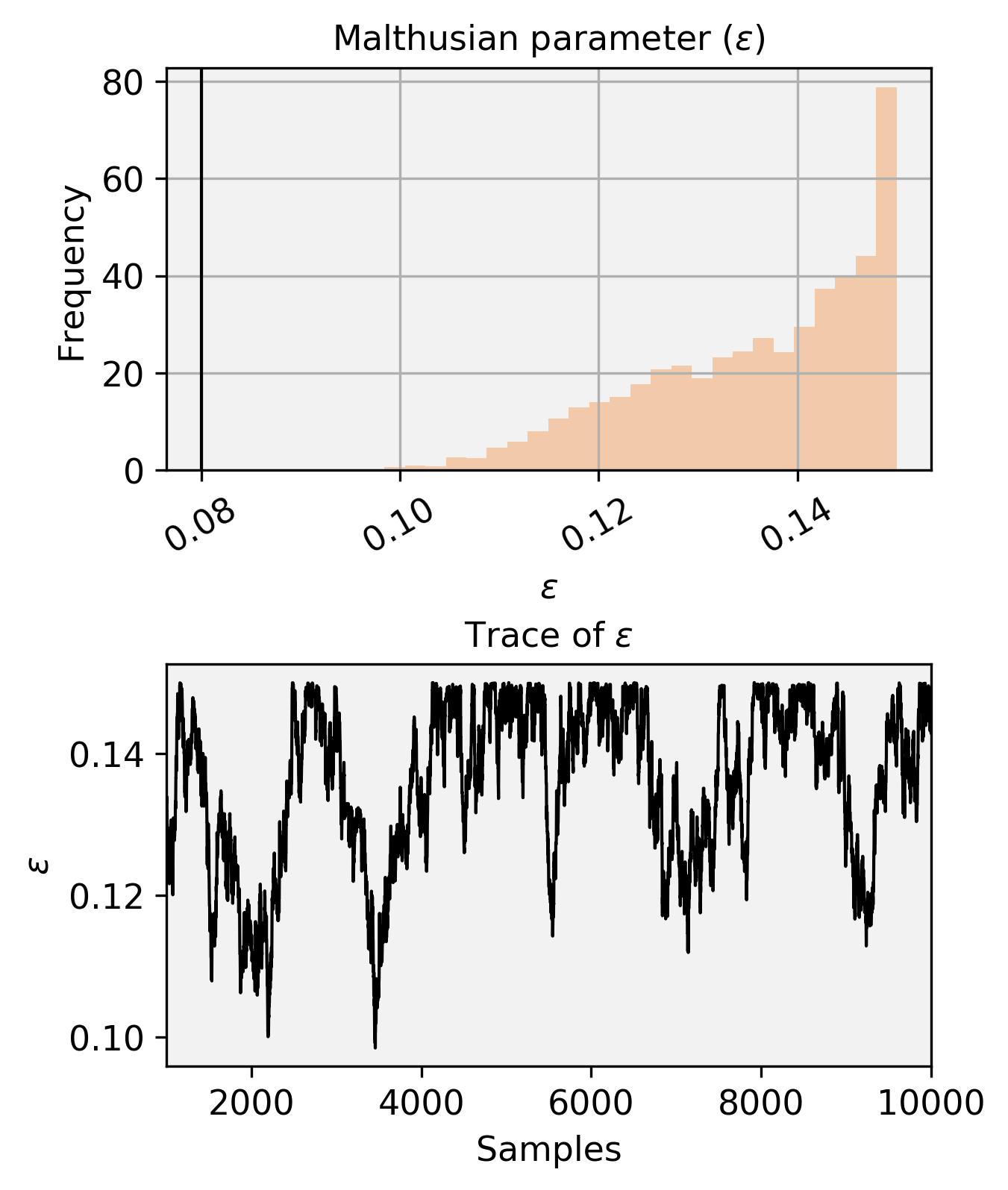}\label{fig:2p-d-malth}}
  \subfloat[]{\includegraphics[width=66mm,valign=t]{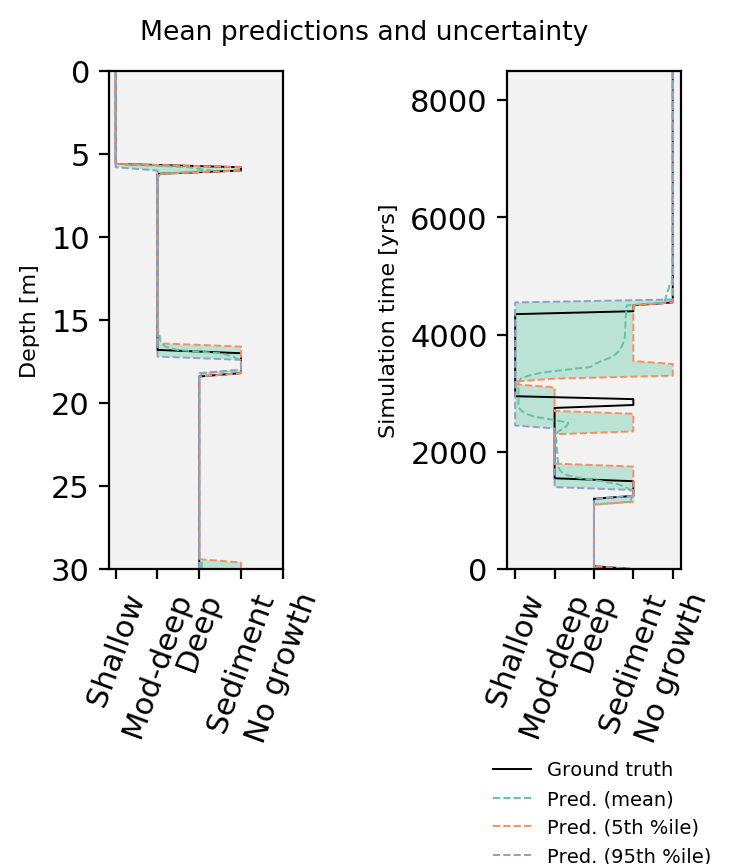}\label{fig:2p-d-mcmcres}}
\caption{ The results  from experiment with only two free parameters, $\alpha_s$ and $\varepsilon$. The time-based likelihood predictions are in Panels~(a) to (c), and depth-based likelihood predictions are in Panels~(d) to (f). Panels~(a,b) and (d,e) give  posterior distributions (upper) and trace plots (lower) for the free parameter. The solid, black line in the histograms indicates the true values used to generate the ground truth. Panels~(c) and (f) are \textit{pyReef-Core}  coralgal assemblage (reef-core) prediction on the basis of the (c)~time structure and (f)~depth structure of the ground-truth.   We compare the credible interval (5th and 95th percentile) and mean prediction  with the ground-truth (black line).}
\label{fig:2p-res}
\end{figure*}

\begin{figure*}[htb!]
\centering
  \subfloat[]{\includegraphics[width=90mm]{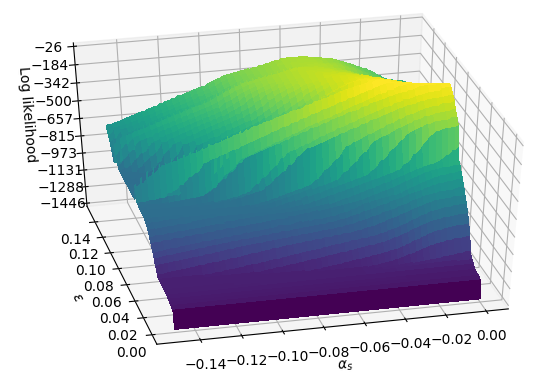}\label{fig:2p-3d-t}}
  \subfloat[]{\includegraphics[width=90mm]{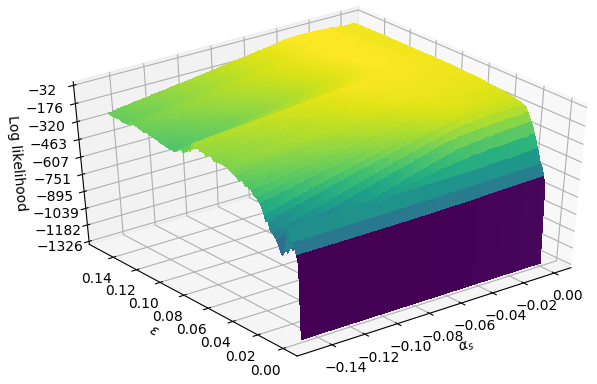}\label{fig:2p-3d-d}}
\caption{Log of the likelihood surface as a function of $\alpha_s$ and $\varepsilon$ using (a)~time-based and (b)~depth-based likelihood.}
\label{fig:2p-3d}
\end{figure*}

\subsection{Results for two free parameters}
Figure~\ref{fig:2p-res} presents the results of two-parameter experiments where the Panels (a)~to~(c) present the results for the time-based likelihood while Panels (d)~to~(f) present the results for the depth-based likelihood. 

We observe that with time-based likelihood the parameter estimates are highly accurate. The posterior distributions of $\alpha_s$ and $\varepsilon$ contain the true values, with the mean and modes nearly centered on the true values (Figures~\ref{fig:2p-t-ay} and \ref{fig:2p-t-malth}; Table~\ref{tab:summ-stats}). In contrast, the posterior distributions of $\alpha_s$ and $\varepsilon$ using  depth-based likelihood  (Figures~\ref{fig:2p-d-ay} and~\ref{fig:2p-d-malth}) are  wider, reflecting  greater uncertainty and in the case of $\varepsilon$. It is unlikely that the histogram estimates approximates the exact posterior distribution    (Figure ~\ref{fig:2p-d-malth}). \textcolor{black}{ The trace plots suggest that the MCMC chain has   converged to a sub-optimal mode  of  the posterior distribution for  Malthusian parameter  $\varepsilon$. This can be seen in Figure \ref{fig:2p-3d}, Panel (b) where a series  of sub-optimal modes are present in the log-likelihood surface.  }

The mean prediction with  the $5\%$ and $95\%$ credible interval for the time-based and depth-based estimation are presented in Figures~\ref{fig:2p-t-mcmcres} and \ref{fig:2p-d-mcmcres} respectively.  The predictions using the time-based likelihood are remarkably accurate at estimating both the time and depth structures of the reef-core data. In contrast, while predictions of the depth structure are accurate, predictions of the time structure are not at certain segmentation times (Panel~f). The prediction shows that the shallow assemblage developed earlier and ceased growth earlier in time, with a longer period of sedimentation between $\sim$3000-4500 years of the simulation. This verifies the concept that many different time-based structures can lead to very similar depth-based structures.

In order to visualise the true log-likelihood, we use a grid-search that considers the combination of selected  parameters  as shown in Figure~\ref{fig:2p-3d}, which  displays the time-based  (Panel~ a)  and depth-based (Panel~b) log-likelihood surface as a function of $\alpha_s$ and $\varepsilon$ . Figure~\ref{fig:2p-3d-t} shows one distinct peak that is centered near the true values of $\alpha_s$ and $\varepsilon$ (-0.03 and 0.08 respectively). In contrast, Figure~\ref{fig:2p-3d-d} shows a flat log-likelihood surface indicating that an area of equally likely combinations of  $\alpha_s$ and $\varepsilon$.

\begin{figure*}[htbp!]
\begin{tabular}{cc}
\subfloat[]{\includegraphics[width=65mm]{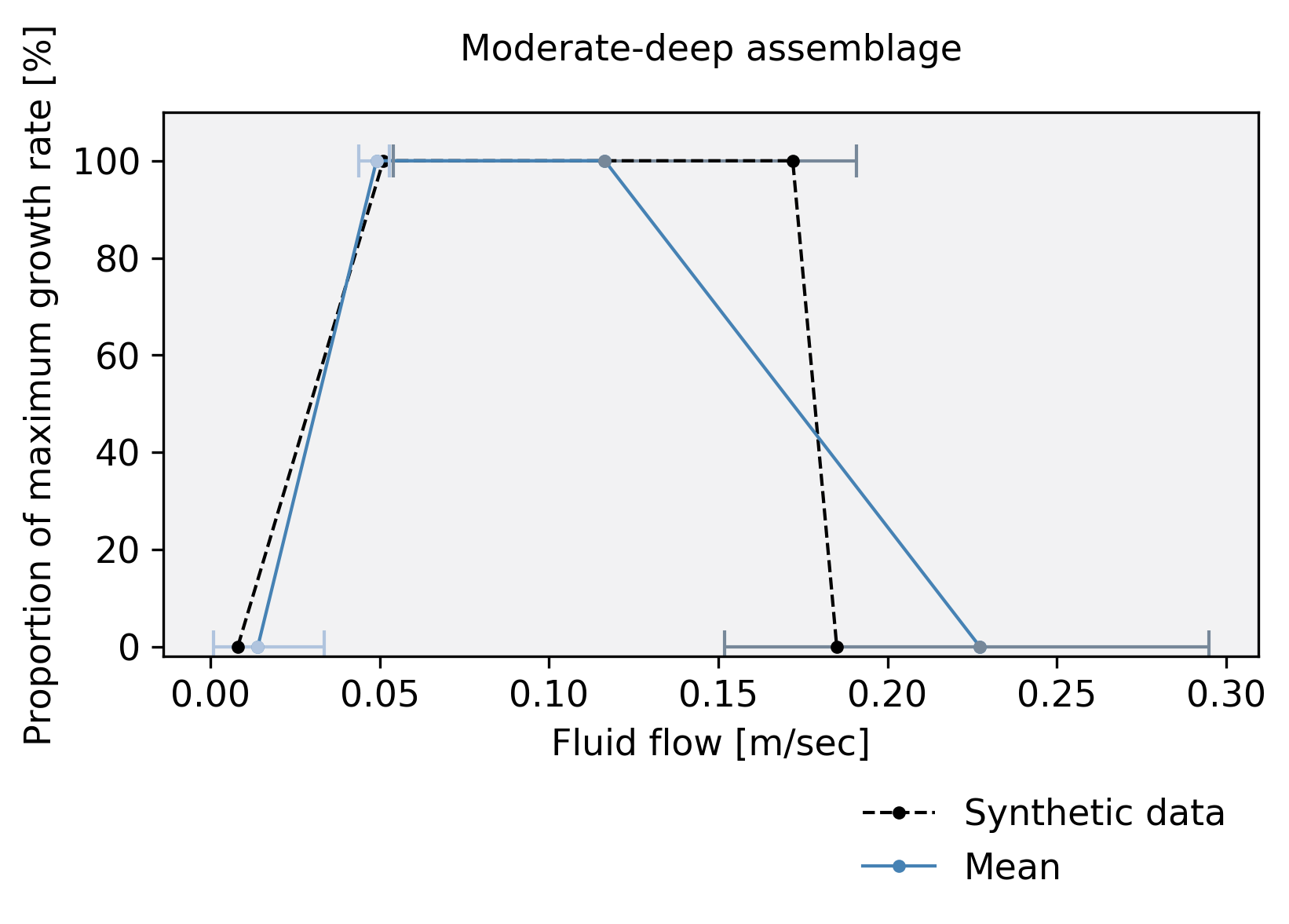}\label{fig:4p-t-flow}} &
\subfloat[]{\includegraphics[width=65mm]{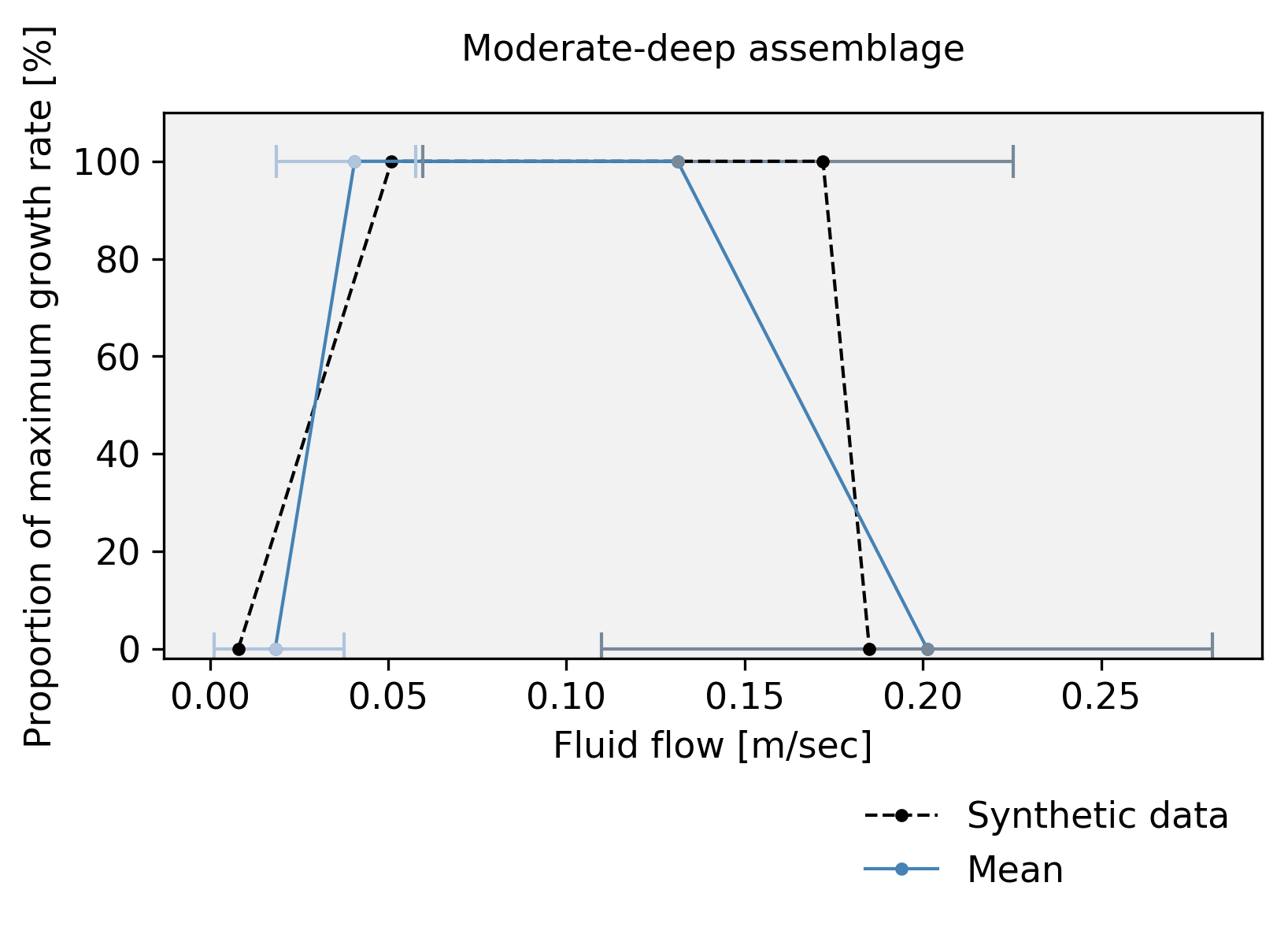}\label{fig:4p-d-flow}}\\
\subfloat[]{\includegraphics[width=90mm]{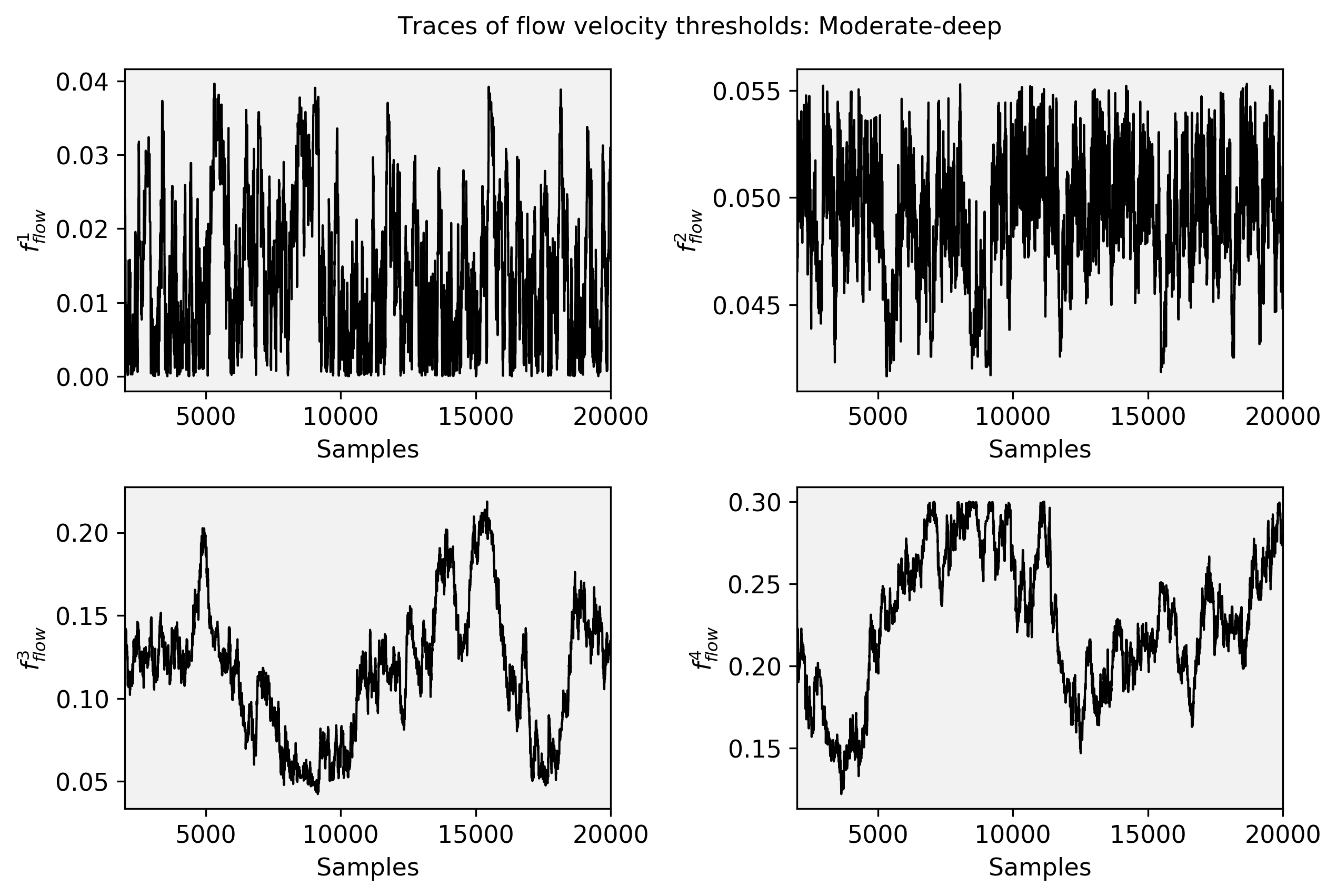}\label{fig:4p-t-trace}} &
\subfloat[]{\includegraphics[width=90mm]{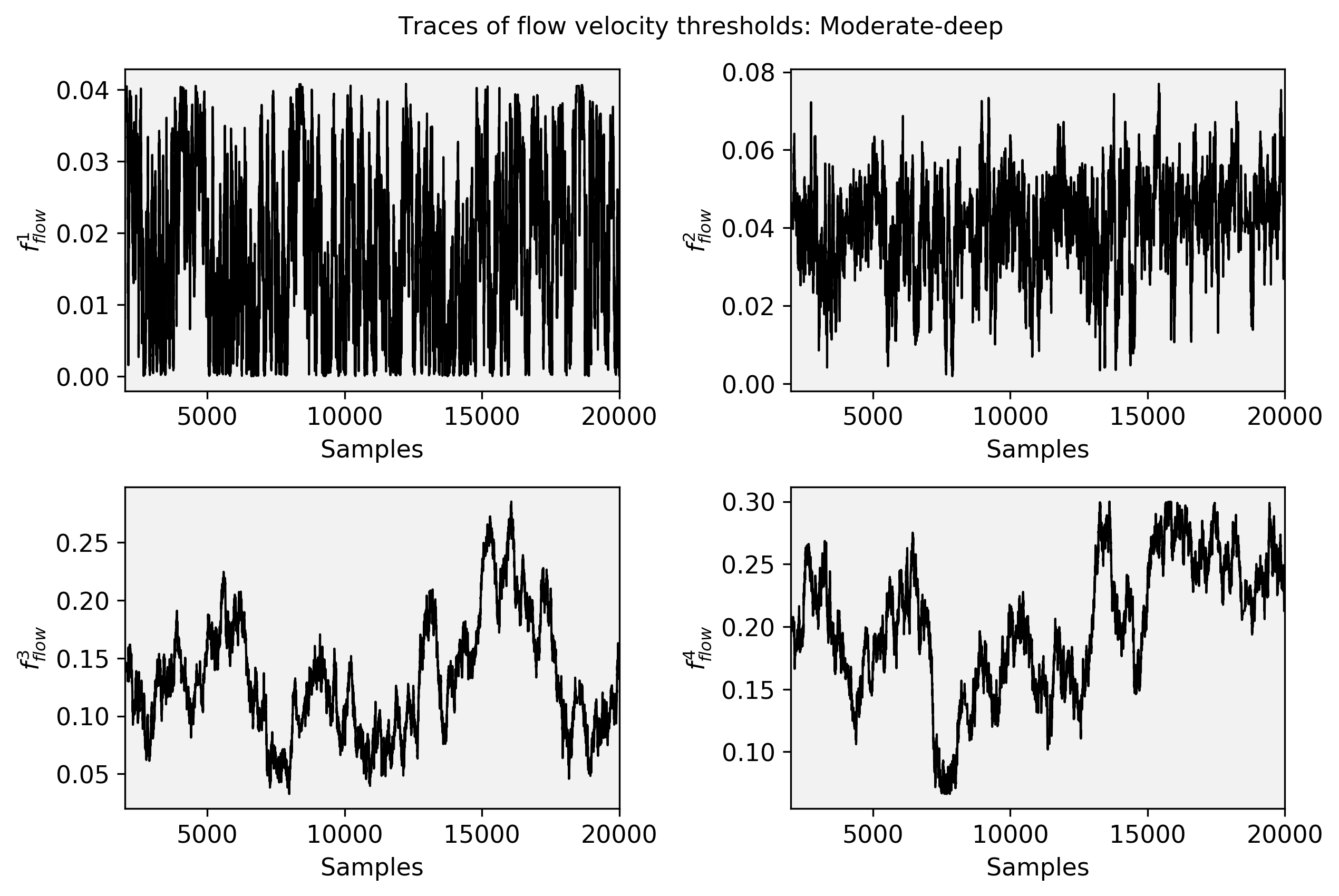}\label{fig:4p-d-trace}}\\
\subfloat[]{\includegraphics[width=70mm]{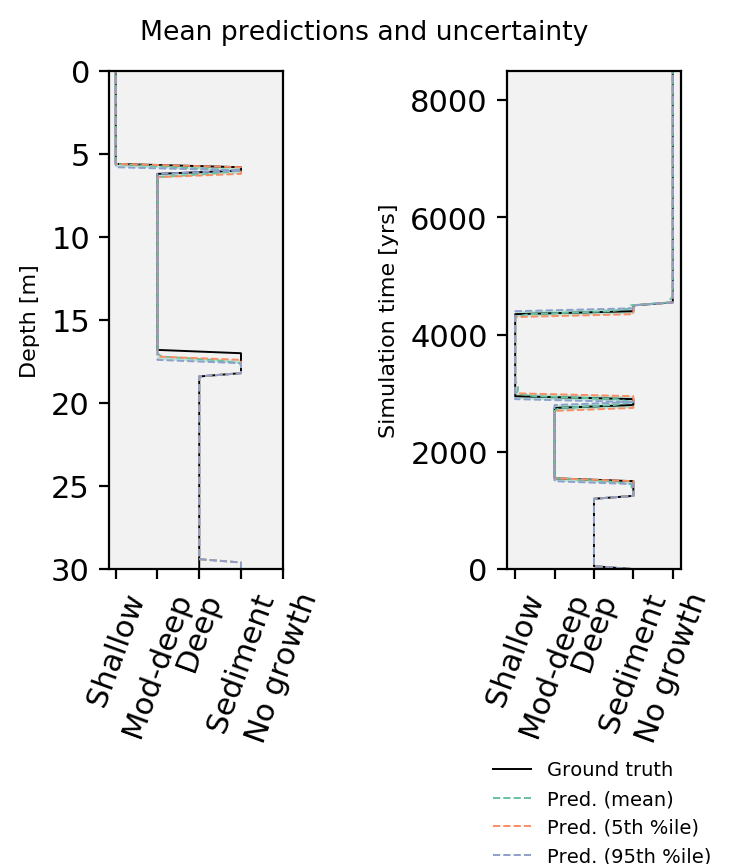}\label{fig:4p-t-mcmcres}} &
\subfloat[]{\includegraphics[width=70mm]{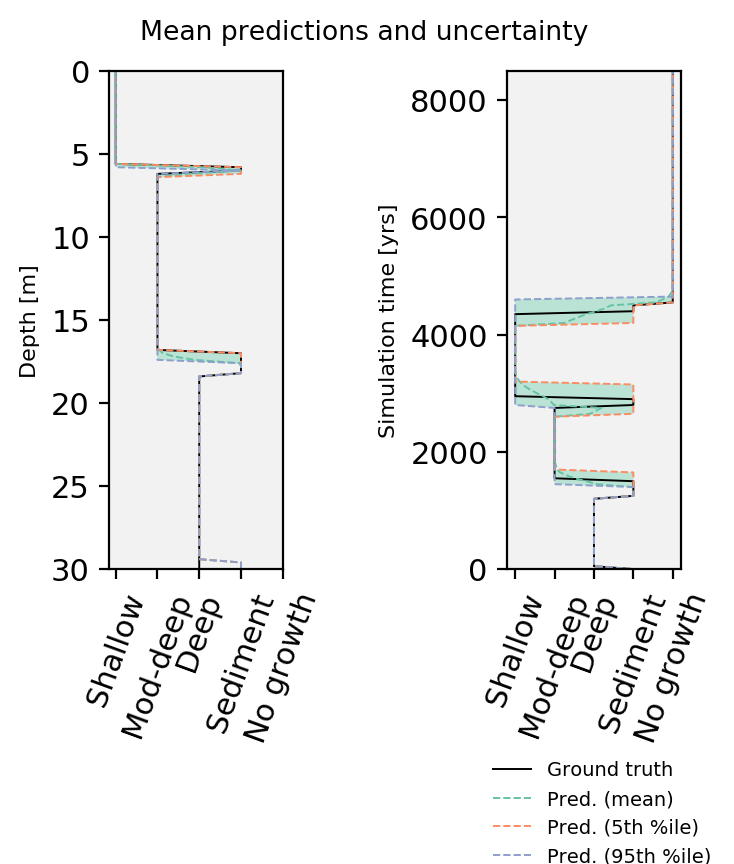}\label{fig:4p-d-mcmcres}}\\
\end{tabular}
\caption{Results for an experiment with four free parameters governing the flow velocity of the moderate-deep assemblage. Panels~(a), (c) and (e) use a time-dependent likelihood and Panels~(b), (d), and (f) use a depth-dependent likelihood. Panels~(a) and (b) are the flow velocity exposure threshold composed of four parameters ($f_{flow}^1,...,f_{flow}^4$, left-to-right), which function as coordinates. They are visualised to represent how coral growth is limited according to the shape of these functions. The black line represents the initial thresholds used to create the synthetic data. The blue line presents the modal estimates of each parameter with an envelope that represents the 95\% credible interval. We show the associated trace-plots of MCMC chains for $f_{flow}^1,...,f_{flow}^4$  in Panels~(c) and (d). Panels~(e) and (f) show  the coralgal assemblage predictions showing the mean and 95\% credible interval .}
\label{fig:4p-res}
\end{figure*}

\begin{figure*}[htb!]
  \subfloat[]{\includegraphics[width=90mm]{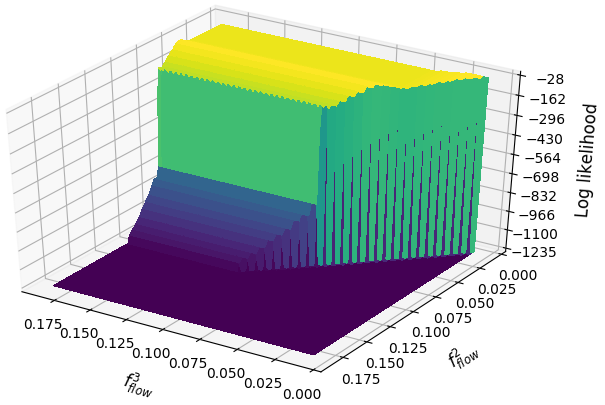}\label{fig:4p-3d-t-23}}
  \subfloat[]{\includegraphics[width=90mm]{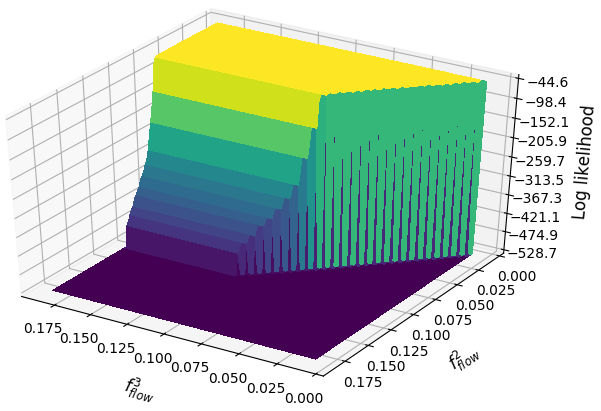}\label{fig:4p-3d-d-23}}\\
\caption{The log-likelihood surface for $f_{flow}^2$ and $f_{flow}^3$ for the moderate-deep assemblage using(a)~time-based  and(b)~depth-based likelihood. The areas of lowest likelihood below the diagonal represent impossible values due to the condition that $f_{flow}^3 \geq f_{flow}^2$ for all values of $f_{flow}^2$.}
\label{fig:4p-3d}
\end{figure*}

\subsection{Results for higher number of free parameters}

We present further results with four free parameters using single-chain MCMC sampling (Algorithm 2) to handle a slightly higher-dimensional problem using random-walk proposal distribution with bi-level constraints shown in Algorithm \ref{alg:cons}. In this setting, we sample the four flow velocity parameters of the medium-deep assemblage, while we fix the remaining parameters to their true values. We summarize the results as shown in  Figure~\ref{fig:4p-res}, where Panel~(a) shows that the parameters $f^1_{2,flow}$ and $f^2_{2,flow}$ are well estimated by the model with time-based likelihood; however, $f^3_{2,flow}$ and $f^4_{2,flow}$ are not well estimated which indicates local convergence.  The trace plots for $f^3_{2,flow}$ and $f^4_{2,flow}$ in Figure~\ref{fig:4p-res} Panels~(c and d),  show a high degree of auto-correlation in the iterates and suggests that the MCMC sampler has converged to one of the sub-optimal modes in the multimodal likelihood landscape. Note that rather than using a histogram, visualise the  the posterior differently using the coordinates of the flow velocity parameters in Figure~\ref{fig:4p-res}, Panel (a and b). We converge to  a sub-optimal mode  since we find that we recover the true value (synthetic data) within the credible interval.     A similar story emerges for the results of depth-based likelihood, where  $f^1_{2,flow}$ and $f^2_{2,flow}$ are well estimated but not as well as the time-based likelihood, while $f^3_{2,flow}$ and $f^4_{2,flow}$ are not that well estimated, and a high degree of auto-correlation for $f^3_{2,flow}$ and $f^4_{2,flow}$ is shown.  

In the sampling and prediction for both time and depth-based likelihood, Figure~\ref{fig:4p-t-flow}), left side of Panels~(e)~and~(f), are generally comparable to the estimates for the two free parameter setting.   We find that the  estimates of the time-based likelihood  are well approximated and  consistent with the two free parameter setting, but not so well captured by the depth-based likelihood as shown in  Figure~\ref{fig:4p-t-flow}), right side of Panels~(e)~and~(f).

We show the log-likelihood surface as a function of $f_{flow}^2$ and $f_{flow}^3$  in Figure~\ref{fig:4p-3d}; Panel~(a) for the time-based likelihood and Panel~(b) for depth-based likelihood. Panel~(a) shows a marked peak for $f_{flow}^2=0.05$; however,  this log-likelihood value is the same across a range of values for $f^3_{flow}$, shown by the ridge at $f_{flow}^2=0.05$.

Furthermore, we evaluate the performance of Bayesreef      using   random-walk (RW) and adaptive random-walk (ARW) proposal distributions.   We show the results in Table \ref{tab:finalresults}  with miss-classification score (MC Score) for reef-core predictions using the depth-based likelihood, for only 3 parameters, that includes the Malthusian and two community interaction parameters ($\varepsilon$,$\alpha_m$,$\alpha_s$). We note that this is because the ARW is not feasible for the rest of the 24 parameters   ($\mb f_{flow_{a1,a2,a3}}$,$\mb f_{sed_{a1,a2,a3}}$) due to the bi-level constraints given in Algorithm 1.  

\begin{table*}[htb!]
\begin{tabular}{l l l l l l }
\hline
 MCMC Method & Parameters &  Parameter List   &   Samples  & MC Score  (mean, std)  & Accepted \% \\
\hline 
RW  & 3 &  [$\varepsilon$,$\alpha_m$,$\alpha_s$] & 5 000 & (2.47, 0.85) & 34.36 \\ 
ARW  & 3 &  [$\varepsilon$,$\alpha_m$,$\alpha_s$] & 5 000  & (2.76,0.69) & 0.86 \\
RW  & 3 &  [$\varepsilon$,$\alpha_m$,$\alpha_s$] & 10 000 & (2.48, 0.88) & 38.56 \\ 
ARW  & 3 &  [$\varepsilon$,$\alpha_m$,$\alpha_s$] & 10 000  & (2.30,0.79) & 0.77 \\
RW  & 3 & [$\varepsilon$,$\alpha_m$,$\alpha_s$]& 20 000 & (2.51, 0.83)& 38.46 \\ 
ARW  & 3 & [$\varepsilon$,$\alpha_m$,$\alpha_s$] & 20 000  &  (2.55, 1.07)   &  1.02  \\

\hline
\end{tabular} 

\caption{Results showing miss-classification score (MC Score)   using depth-based likelihood for different scenarios.  
}
 
\label{tab:finalresults}
\end{table*}

\begin{figure*}[htb!]
  \subfloat[]{\includegraphics[width=90mm]{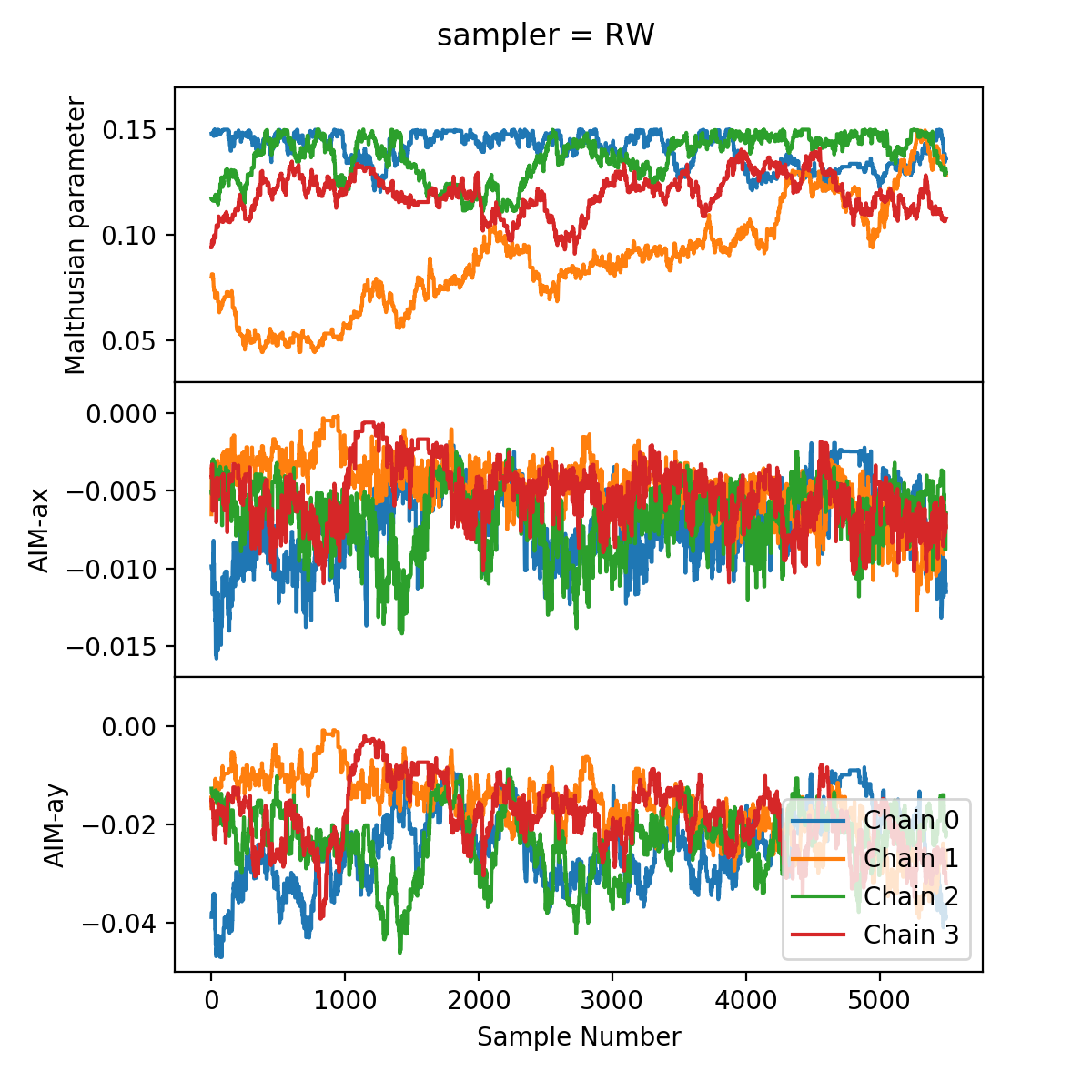}\label{fig:sampler-trace-rw}}
  \subfloat[]{\includegraphics[width=90mm]{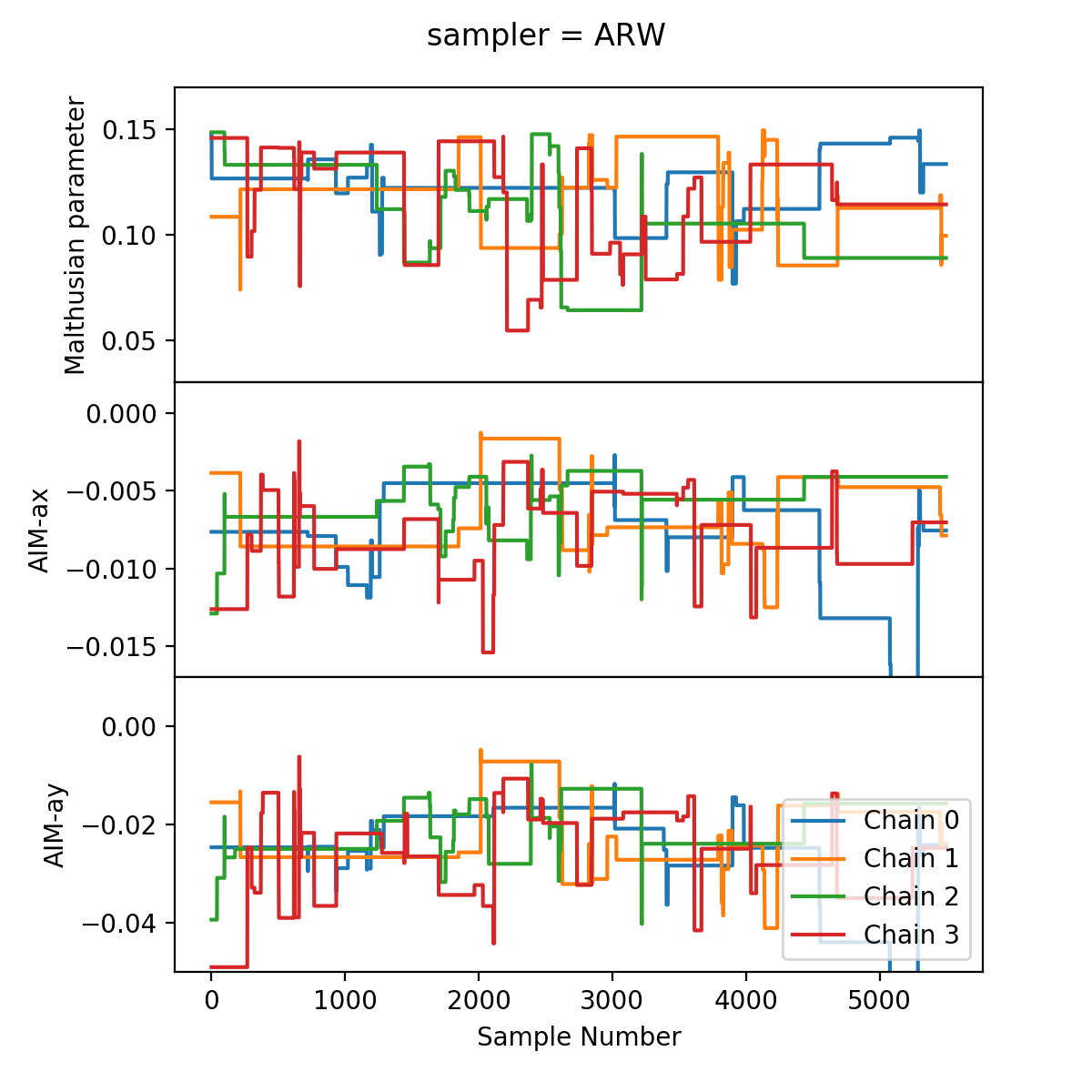}\label{fig:sampler-trace-arw}}\\
\caption{Trace plots for parameters sampled via random-walk (RW) and adaptive random-walk  (ARW) proposal distribution of the single-chain MCMC sampler.  The different colors in the sampling trace-plot represent  runs from four different starting points, after a burn-in of 3000 samples. }
\label{fig:sampler-trace}
\end{figure*}

 As a measure of convergence, we run multiple chains  using different initial values, and use them to calculate the potential scale reduction factor (PSRF) \cite{gelman_inference_1992} for each parameter; this is the ratio of each parameter's sampling variance across all chains to its variance within any single chain.  Values of the PSRF near 1 indicate convergence, whereas values much larger than 1 indicate either significant posterior uncertainty due to limited samples or that individual chains have not yet explored the entire posterior distribution.  Figure \ref{fig:sampler-trace} shows trace plots of the iterates for the Malthusian parameter and for the elements of the competition matrix which gives further insights for the 3 free parameters for 10,000 samples in Table \ref{tab:finalresults}.  Figure~\ref{fig:sampler-trace} shows that the independent chains from both proposal distributions have integrated auto-correlation times for all parameters that are comparable to their total length, making them difficult to estimate numerically.  However, AWR does a better job of making large jumps in parameter space that different sample regions of the posterior, despite its lower overall acceptance rate.   In the Malthusian parameter, we get the PSRF of 6.6  for the RW sampler which is evident in the broadly divergent trace plots seen in the top panel of Figure~\ref{fig:sampler-trace-rw}.  For the ARW sampler,  we get PSRF of only 1.16  for the Malthusian, while  the rest of the parameters have less than 1.2 PSRF.  We get a lower acceptance rate for  ARW when compared to RW proposal distribution and despite this, we find that ARW   provides a  better estimate of the uncertainty in the final results.

\subsection{Parallel tempering MCMC }

\begin{figure}[htbp!]
  \begin{center}
    \begin{tabular}{cc} 
      \subfloat[Posterior distribution (parallel tempering MCMC)]{\includegraphics[width=65mm]{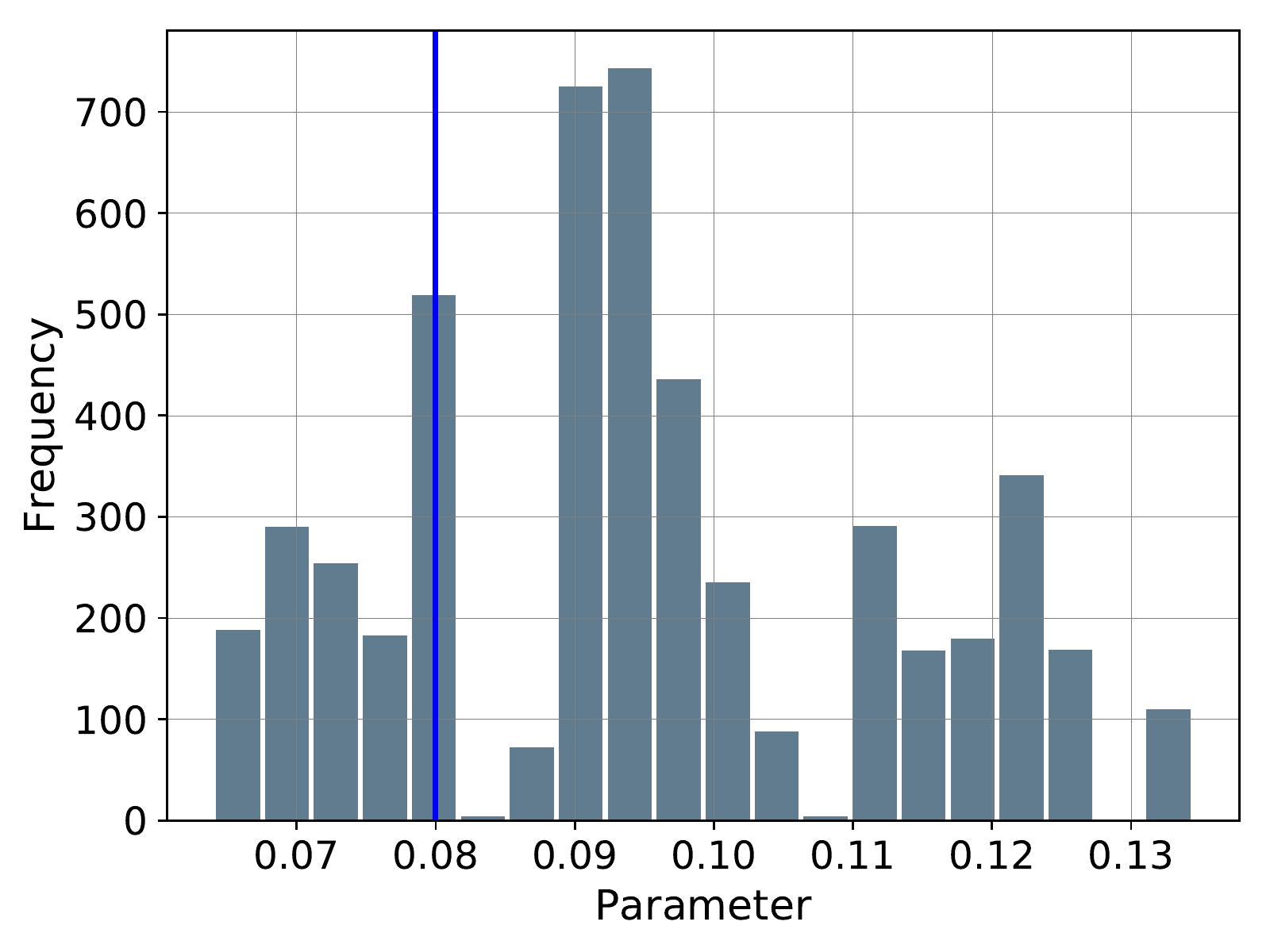}}\\
      \subfloat[Trace-plot - 10 replicas (parallel tempering MCMC)]{\includegraphics[width=65mm]{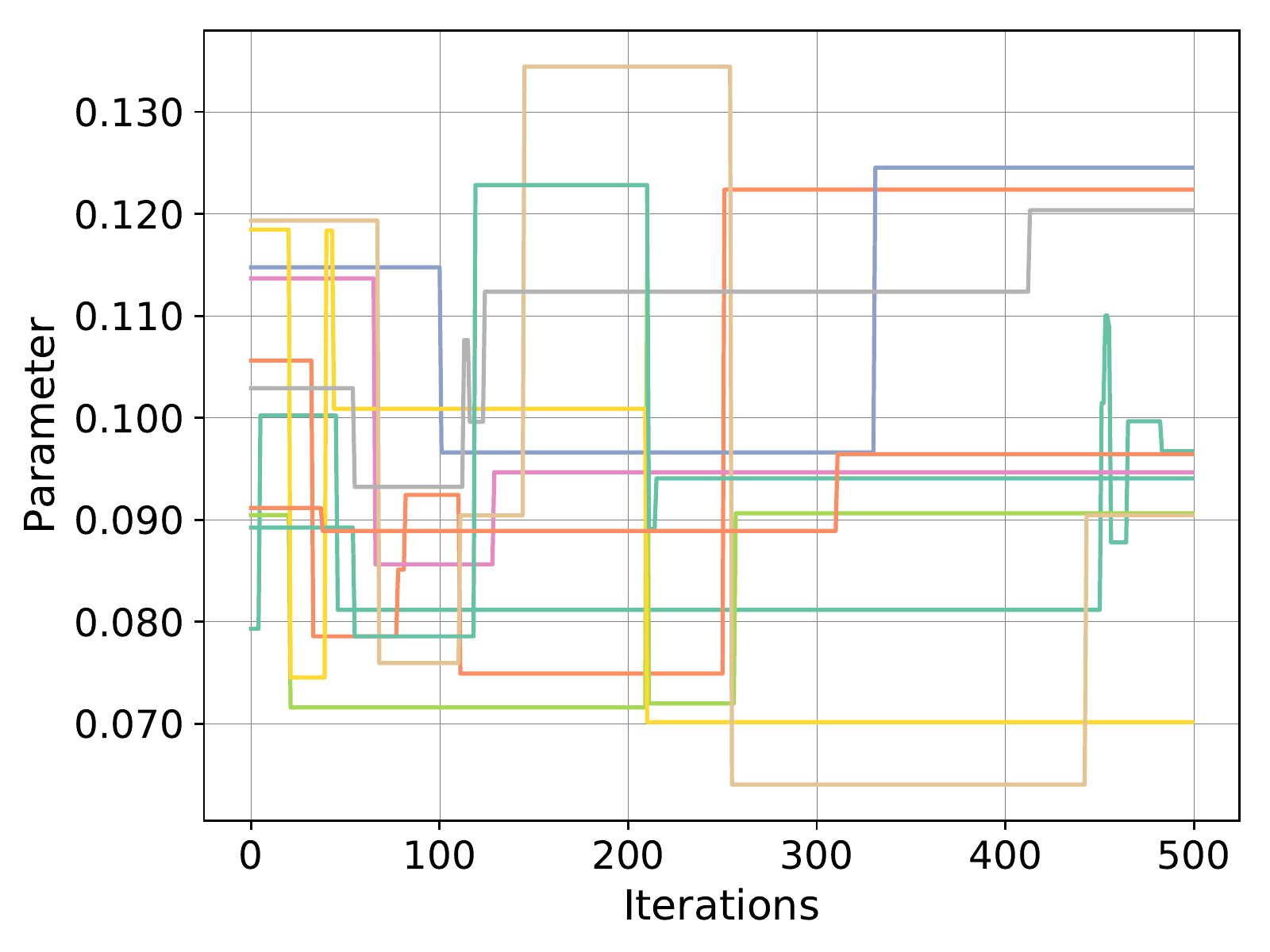}}\\

    \end{tabular}
    \caption{  Example of posterior distribution using single-chain and parallel tempering MCMC  (Malthuian parameter)  for synthetic reef-core experiment using 10,000 iterates. The true value that was used to generate the synthetic reef-core is shown as a vertical  blue line.  }
 \label{fig:syn_pos}
  \end{center}
\end{figure}

\begin{figure}[htb!]
\begin{tabular}{cc}
\subfloat[Single-chain MCMC]{\includegraphics[width=45mm]{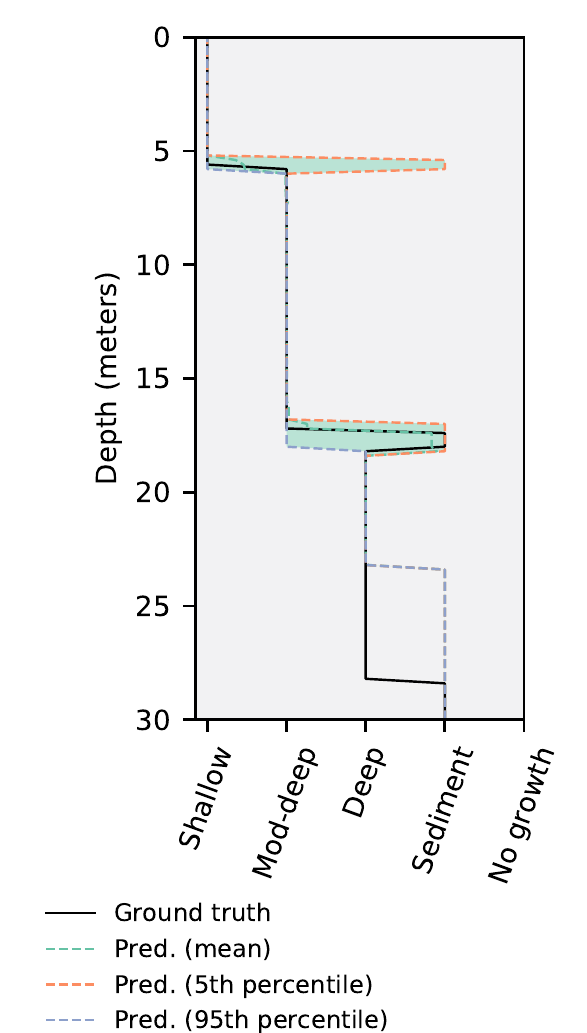} } &
\subfloat[Parallel tempering  MCMC]{\includegraphics[width=45mm]{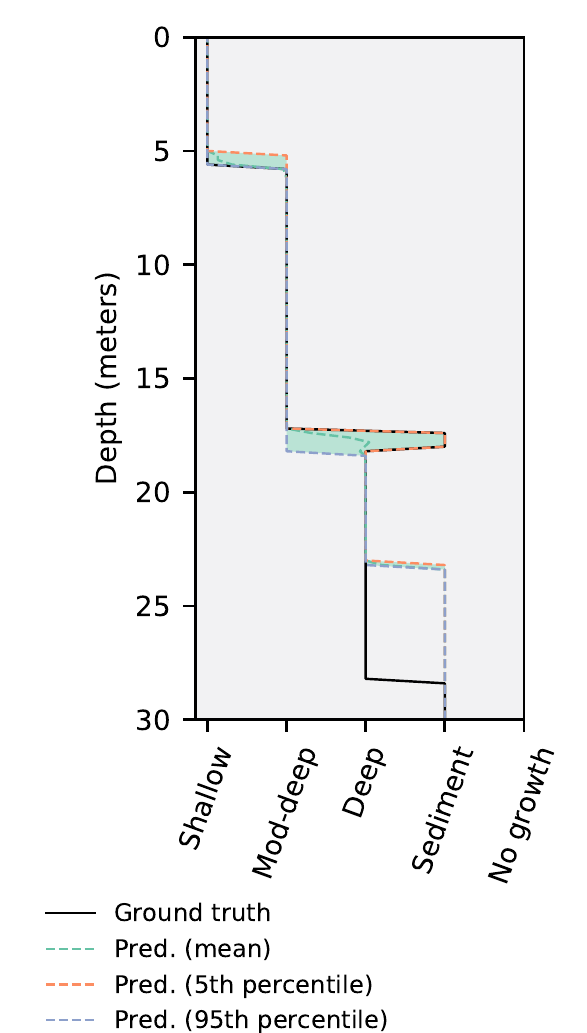} }  
\end{tabular}
\caption{\textcolor{black}{Reef-core predictions using single-chain and parallel tempering MCMC with 20,000 iterates (samples) for the synthetic-core data.}}
\label{fig:syntheticpt}
\end{figure}

\textcolor{black}{Using   depth-based likelihood, we compare the  results of single-chain (SC)   with parallel tempering (PT) MCMC. We use burn-in phase (50 \% ) with 10 replicas and execute the algorithms on single processing cores. Table \ref{tab:ptsyn} shows a summary of results with misclassification (MC) score given by mean and standard deviation, acceptance percentage, and approximate computational time for the different number of samples. We notice that while both methods have similar MC score that falls within the range of the standard deviation, SC-MCMC  has a much lower acceptance rate. This could be due to lack of exploration ability when compared with PT-MCMC that explores multimodal posterior distribution.   Figure \ref{fig:syn_pos} shows the posterior distribution and trace-plot for a selected parameter, where we recover the true value (shown as vertical blue line)   as one of the modes in the multimodal posterior. Figure \ref{fig:syntheticpt} shows the prediction for the reef-core by the respective  methods for 10,000 iterates after removing the burn-in. We notice that both methods have a similar prediction performance in terms of MC rate; however, PT-MCMC shows more uncertainty in prediction at specific intervals.  }

\begin{table*}[htb!]

\begin{tabular}{l l l  l l l }
\hline
  Method & Parameters   &   Samples  & MC Score    & Accept \% & Time (hours) \\
\hline 
 
SC  & 27 &   10 000 & (48.33,14.55) &  1.15 & 36.20 \\ 
PT  & 27 &   10 000  & (51.79, 14.45) & 5.48  & 35.80 \\ 
SC  & 27 &   20 000 & (50.35,12.00) &  0.18 & 68.50\\ 
PT  & 27 &   20 000  & (56.26,13.45) & 1.36  & 69.20  \\ 
 
\hline
\end{tabular} 

\caption{\textcolor{black}{Results showing miss-classification score (MC Score) with (mean, standard deviation)   using depth-based likelihood for different methods for the synthetic reef-core, and the predictions are shown in Figure \ref{fig:syntheticpt}. 
}}

\label{tab:ptsyn}
 
\label{tab:ptmcmc}
\end{table*}

\subsection{Application to  reef-core from the Great Barrier Reef}

\begin{figure}[ht!]
\centering
\includegraphics[width=60mm]{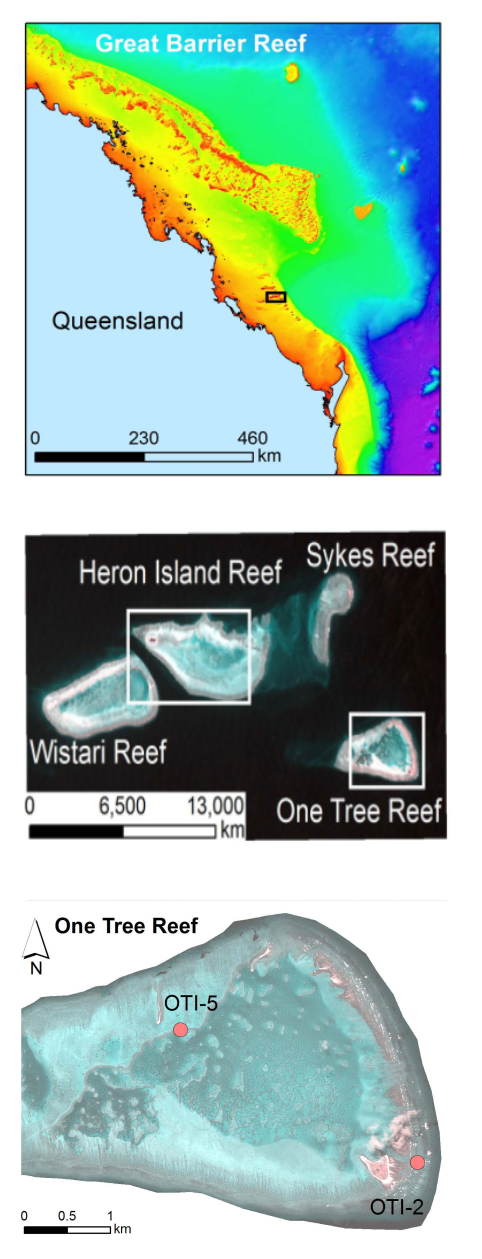}
\caption{\textcolor{black}{Reef-core was drilled  from the Capricorn-Bunker Group of the Southern  Great Barrier Reef, One Tree Reef island  (OTI-5) \cite{salas2018holocene}.} }
\label{fig:heron}
\end{figure}

 We selected a fossil reef-core  from the Capricorn-Bunker Group of the Southern Great Barrier Reef (GBR), occurring 80 kilometers from the Queensland Coast. This is  among one of the most studied location in the GBR due to the presence of a research station that has hosted decades of biological and geological reef research    (Figure  \ref{fig:heron}). In our experiments, we selected OTI-5  that was collected from the northwestern
leeward, low-energy margin   and records a  complex history of fossil reef changes \cite{dechnik2015holocene}. We find that  the modern coralgal assemblages found on OTR  match well with fossil coralgal
assemblages   classified by \cite{dechnik2015holocene, dechnik2016evolution} in terms of
coralgal composition, location and palaeo-water depth indicators.

\begin{figure}[ht!]
\centering
\includegraphics[width=60mm]{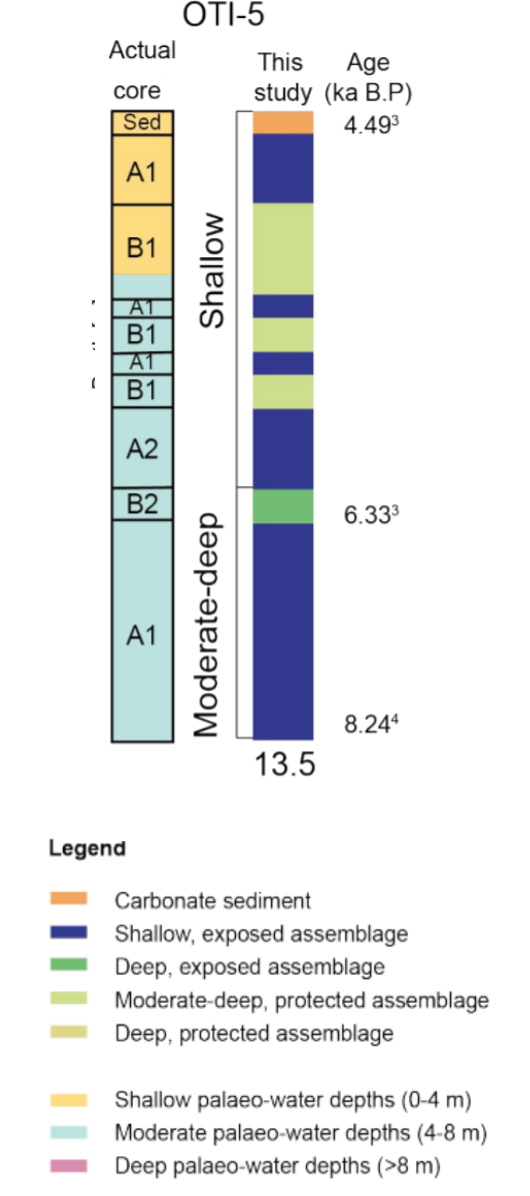}
\caption{\textcolor{black}{Selected OTI-5 from Figure \ref{fig:heron}, we show details of the composition of the reef-core   given six  different types of assemblages.}}
\label{fig:heron_core}
\end{figure}

 While data exists for the assemblages in the  reef-core (i.e. characteristics, depths, ages), the
response of these assemblages to prevailing environmental conditions can only be inferred
indirectly from data, and are not systematically quantified. Figure  \ref{fig:heron_core} shows the details of the composition of the reef-core given different types of assemblages and assemblage age information.   \textcolor{black}{We note that unlike that synthetic reef-core experiments that feature three assemblages, the real reef-core features six assemblages; where three are in the windward (W) and the other three are in the leeward (L) side. This requires an estimation of 6 x 4 flow-velocity parameters and 6 x 4 sediment-input parameters, along with 2 AIM and 1 Malthusian parameter. Hence, we have 51 parameters in total that are estimated by Bayesreef, which presents a more difficult situation. }
 
\begin{figure}[htb!]
\begin{tabular}{cc}
\subfloat[20 000 iterates]{\includegraphics[width=45mm]{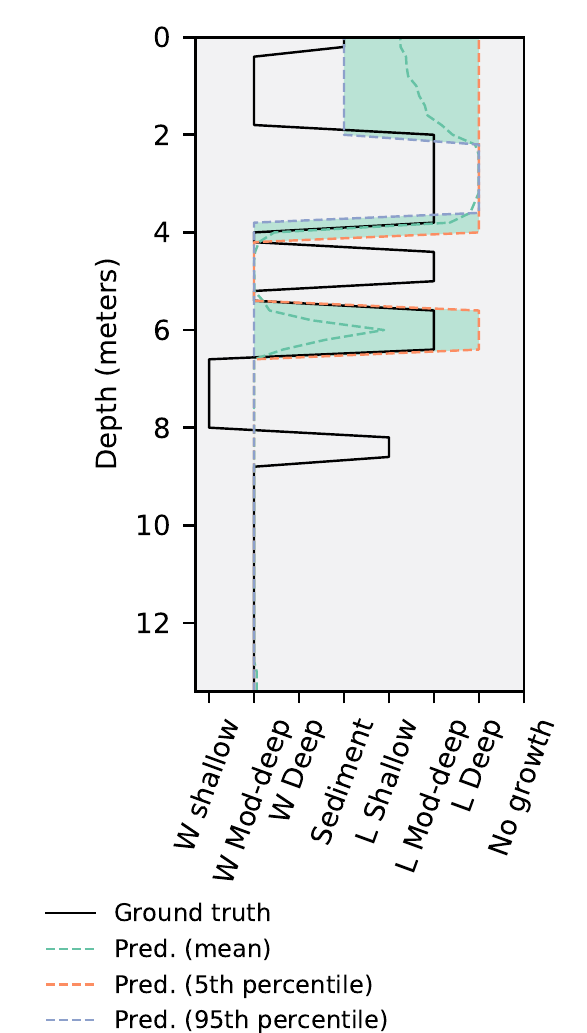} }  
\subfloat[10 000 iterates]{\includegraphics[width=45mm]{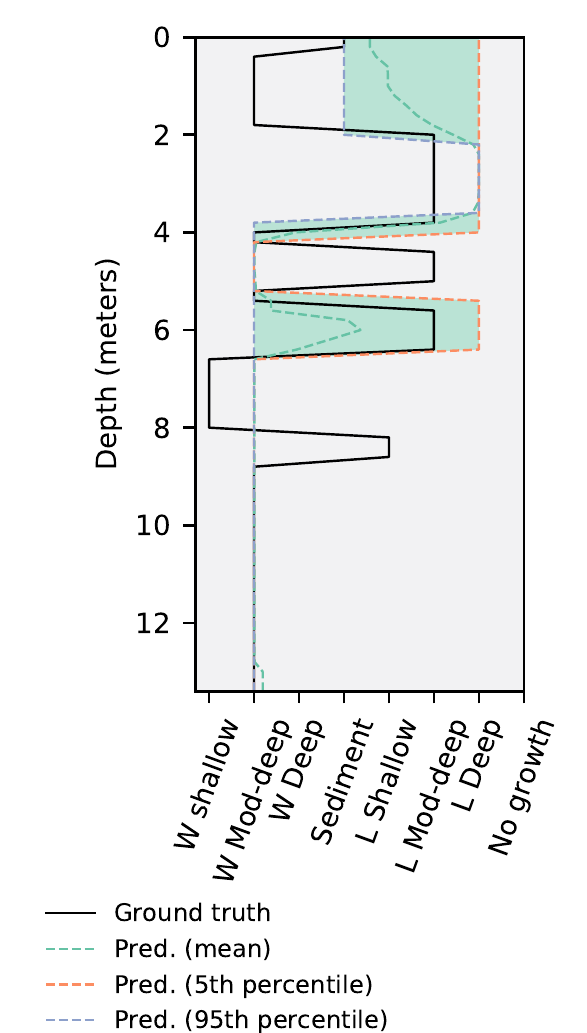} }    
\end{tabular}
\caption{\textcolor{black}{Reef-core predictions  using parallel tempering MCMC for One Tree Island reef-core (OTI-5) ground-truth  given in Figure \ref{fig:heron_core}.  }}
\label{fig:henonpredictions}
\end{figure}

\textcolor{black}{The results in Figure \ref{fig:henonpredictions} show that Bayesreef can successfully predict the windward (W) shallow assemblages and the sediment deposition, but has not been able to predict the deep assemblages very well. However, the predictions are close given the uncertainty quantification for the respective experiments. This could be due to the limitations in the proposal distribution and challenges of sampling given the irregular multimodal posterior distributions shown earlier.   Table \ref{tab:reshenon} gives further details for each experiment configuration, where we see that the core predictions are similar given the MC score. OTI-5
displays a complex reef growth history \cite{dechnik2016evolution}, rather than a
typical shallowing-upward sequence. Therefore the mismatch
between the model predictions and actual core data may
indicate that our prior assumptions about  
some environmental parameters through time (i.e.
sediment input and flow velocity), and their thresholds are currently too poorly constrained in the  current \textit{pyReef-Core} set-up}.

\begin{figure}[htbp!]
  \begin{center}
    \begin{tabular}{cc} 
      \subfloat[Histogram showing posterior distribution]{\includegraphics[width=65mm]{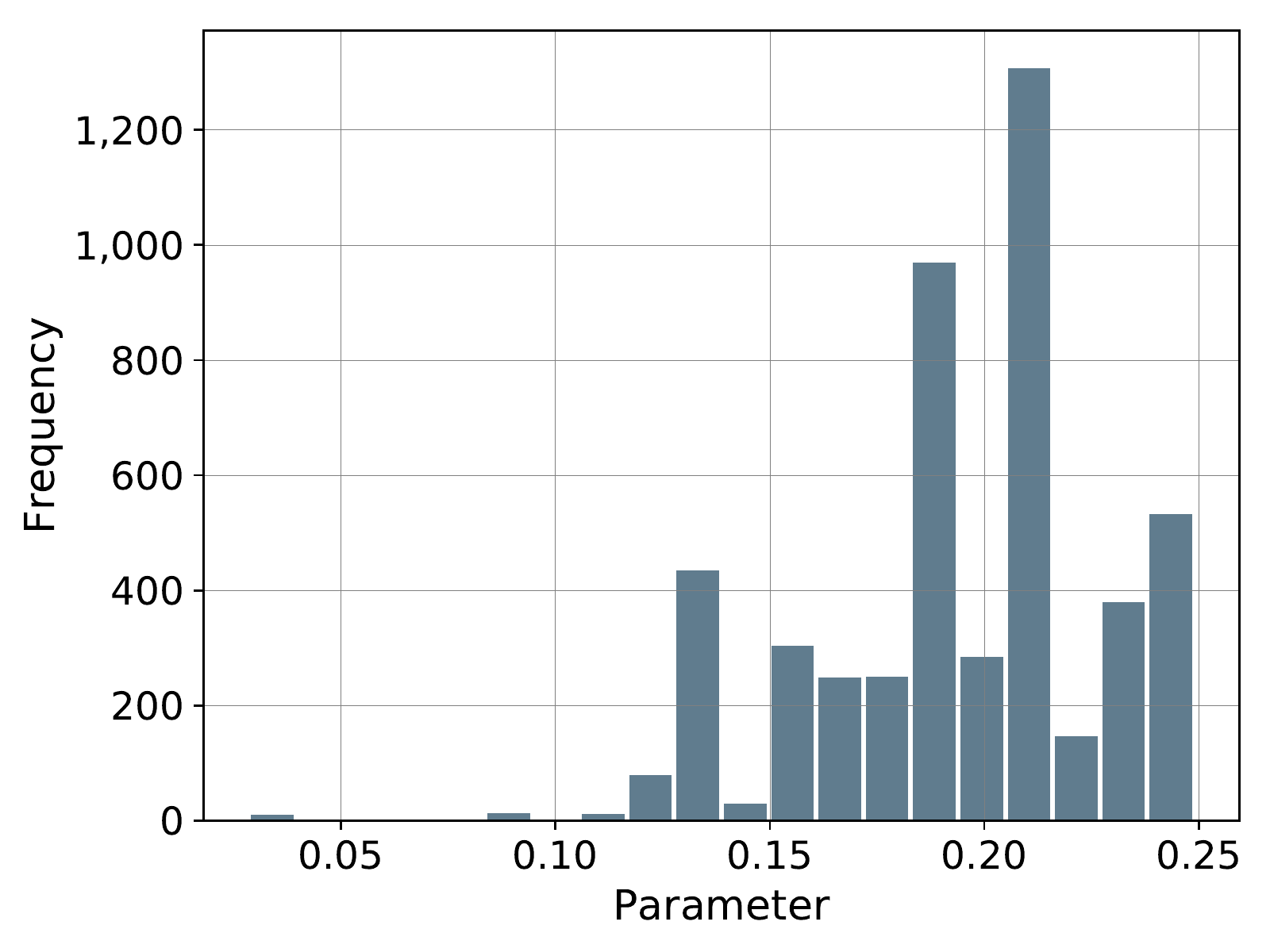}}\\
      \subfloat[Trace-plot of samples for 10 replicas]{\includegraphics[width=65mm]{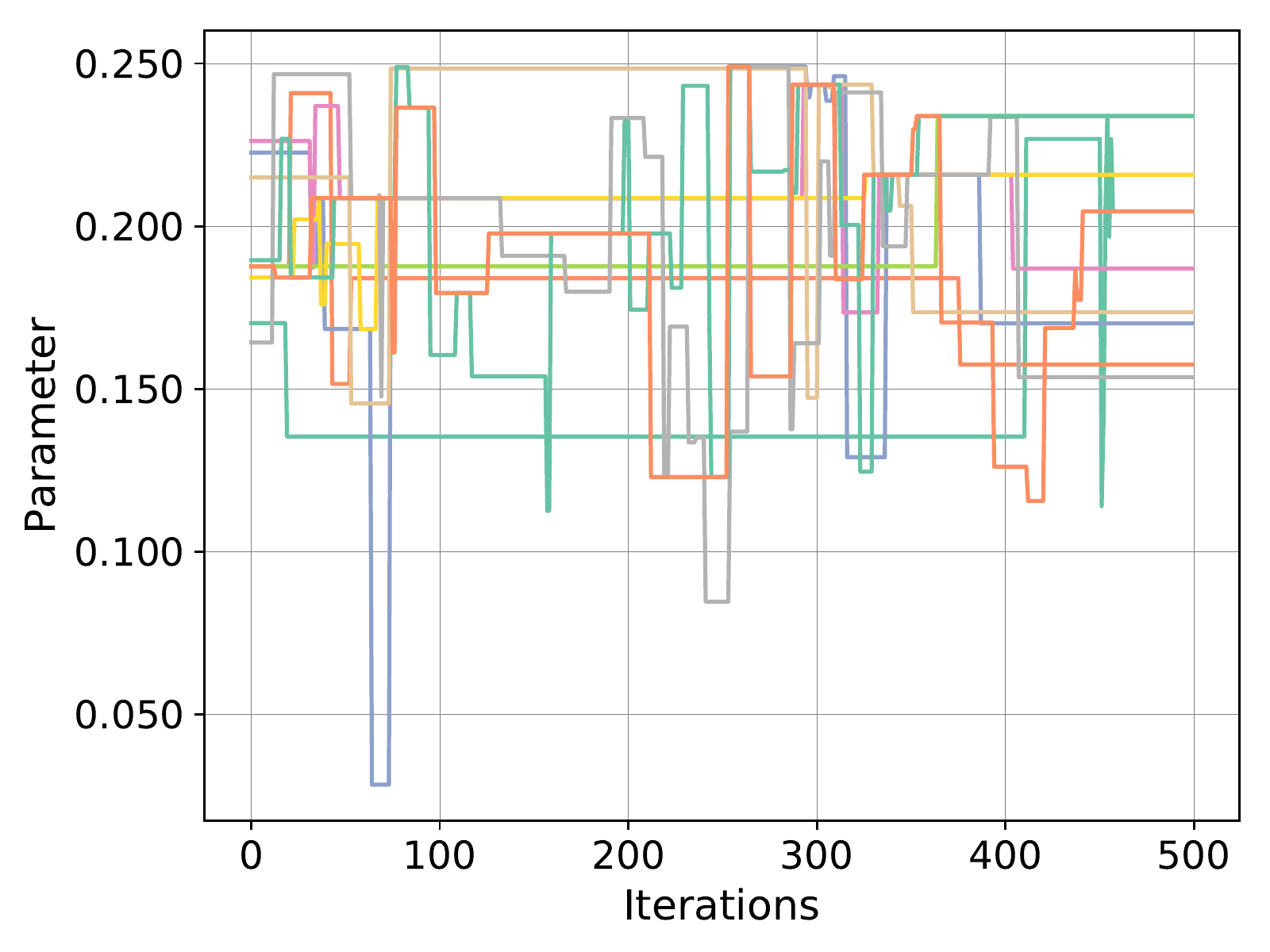}}\\

    \end{tabular}
    \caption{  \textcolor{black}{Example of posterior distribution of Malthusian parameter  using parallel tempering MCMC  for OTI-5 reef-core experiment summarized in Table \ref{tab:reshenon}. }  }
 \label{fig:henon_pos}
  \end{center}
\end{figure}

\textcolor{black}{Figure \ref{fig:henon_pos} shows an example of the posterior distribution of a selected parameter, where we notice that there is multimodality present and with the multiple replicas and there is a good exploration of the parameter space.} 

\begin{table}[htb!]
\begin{tabular}{l l l  l l }
\hline
  Iterates  & Parameters   &   Samples  & MC Score   & Accept \% \\
\hline  
10 000 & 51 & 10 000 & (65.17,15.60) & 25.92 \\
20 000   & 51 &   20 000 & (64.17,
15.79) & 4.46  \\ 
 
\hline

\end{tabular} 

\caption{\textcolor{black}{Results showing miss-classification score (MC Score) with (mean, standard deviation)   using depth-based likelihood with parallel tempering MCMC for selected One Tree Island reef core (OTI-5). The predictions are shown in Figure \ref{fig:heron_core}. We use different configurations (Exp) that considered different limits (priors) for the respective parameters. }
}
 
\label{tab:reshenon}
\label{tab:henon}
\end{table}

\begin{figure}[htbp!]
  \begin{center}
    \begin{tabular}{cc} 
      \subfloat[Flow-velocity parameters]{\includegraphics[width=75mm]{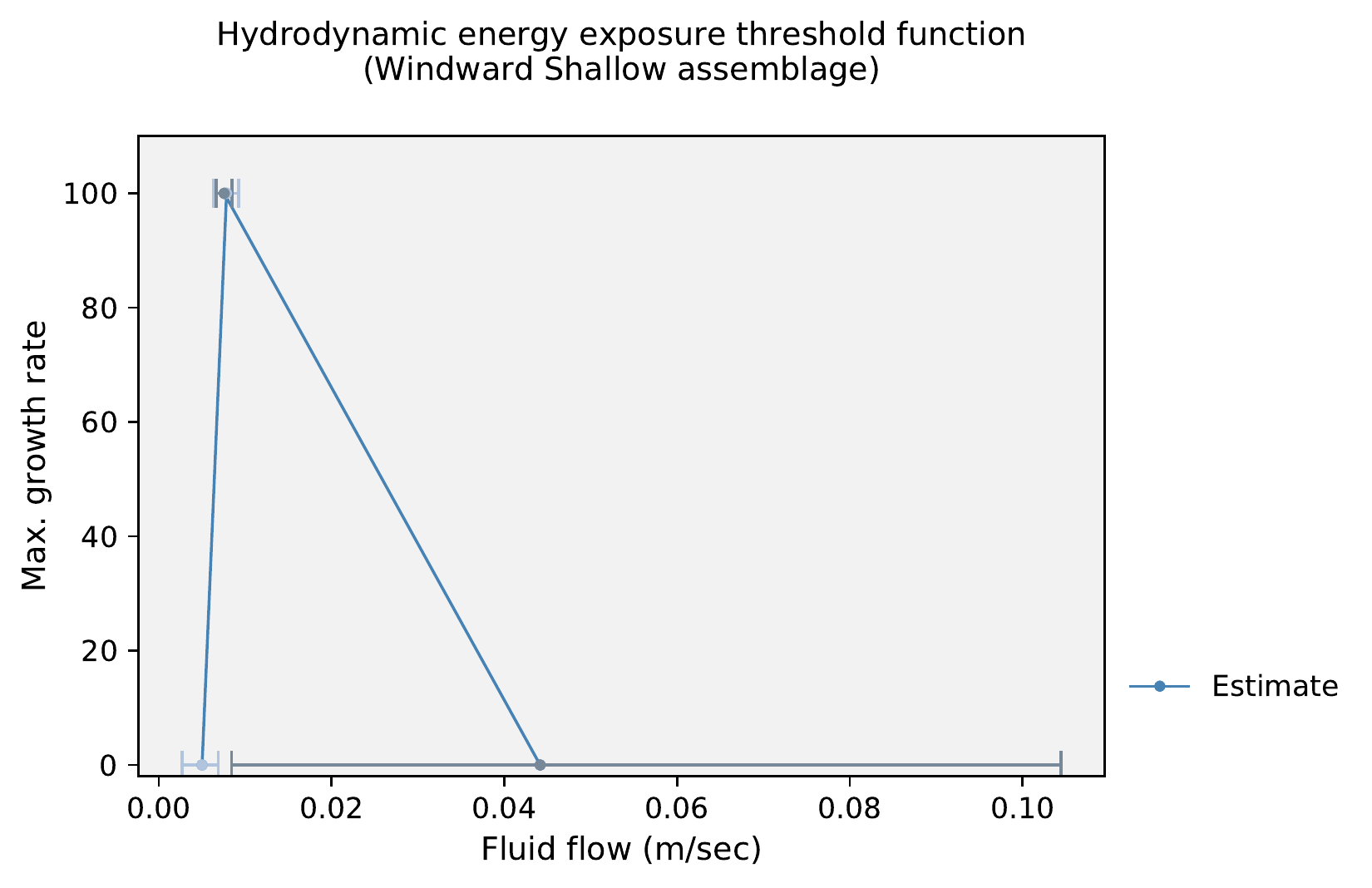}}\\
      \subfloat[Sediment-input parameters]{\includegraphics[width=75mm]{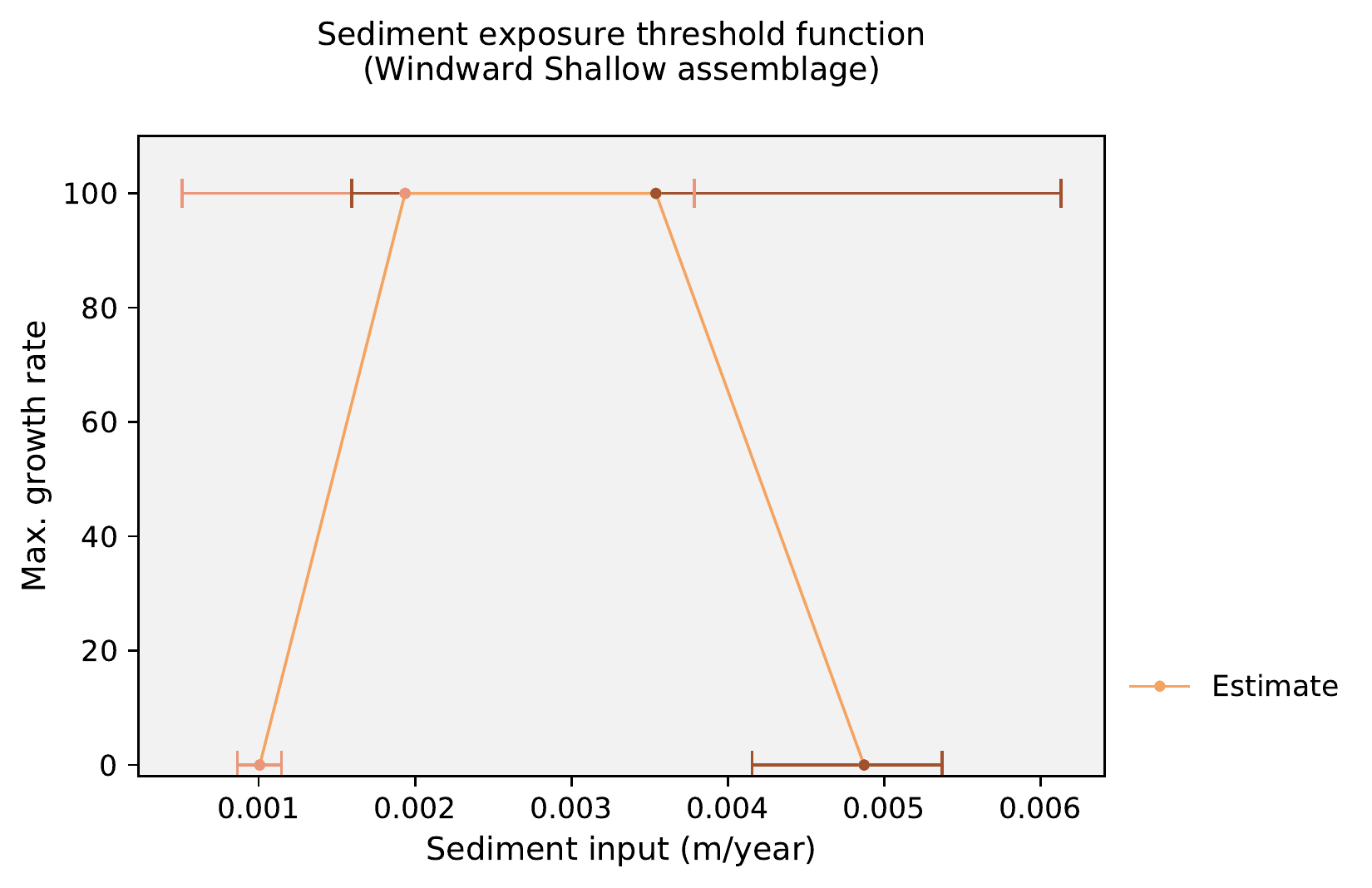}}\\

    \end{tabular}
    \caption{  \textcolor{black}{Example of posterior distribution of flow-velocity and sediment-input parameters  using parallel tempering MCMC  for OTI-5 reef-core experiment summarized in Table \ref{tab:reshenon}. Note that the mean prediction is shown in solid line with  95 \% credible interval as error bar. }  }
 \label{fig:oti_sedflow}
  \end{center}
\end{figure}
 
Figure \ref{fig:oti_sedflow} shows an example of estimation  of flow-velocity and sediment-input parameters for the first assemblage (W-shallow)  using parallel tempering MCMC  for OTI-5 reef-core prediction  shown in Figure \ref{fig:henonpredictions}.

\section{Discussion}

We apply Bayesreef to infer how coralgal community
structure changes in relation to prevailing changes in accommodation, hydrodynamic
energy and sediment input. \textcolor{black}{The results show that \textit{Bayesreef} provides a reliable prediction of the synthetic ground truth data in the experiments that consider the depth and time-based likelihood and different combination of selected free parameters.  Moreover, the estimated parameters provide an accurate prediction of the reef-core when compared to the synthetic reef-core data, which demonstrates convergence to one of the modes of the multimodal posterior distribution}. Bayesreef  provides the groundwork for an insight into the complex posterior distributions of parameters in \textit{pyReef-Core}.

We constrained the number of dimensions in  the experiments to investigate how \textit{Bayesreef} performs with  low-dimensional   and  high  dimensional settings. The  results show that depth-based likelihood gives equally-accurate predictions in selected low  and high-dimensional experiments.  This  promotes confidence in this technique in achieving accurate predictions with higher-dimensional problems.  However, this is limited to synthetic reef-core with a limited number of unknown parameters. Therefore, we need more experiments in future work to evaluate them for the real-world reef-core example. Regardless of the type of likelihood, the experiments could not constrain certain parameters. The large uncertainty around some of the mean estimates  highlight how a broad range of values can achieve the equivalent core prediction (Figures~\ref{fig:4p-t-flow} and \ref{fig:4p-d-flow}). Similarly, the  log-likelihood surface show that, beyond a certain point, any given parameter values produce equivalent high-likelihood values (Figure~\ref{fig:4p-3d}). This indicates that we do not have enough information available to constrain these parameters and informative priors or reef-core data containing more information may be able to improve the inference.

Alternatively, the results could reflect the environmental sensitivity of the moderate-deep assemblages. As the lower limit of fluid flow is better constrained, we may understand moderate-deep coralgal assemblages can tolerate extremely low flow, $\leq$5~centimeters/second (Figure \ref{fig:4p-t-flow}). In addition the lack of constraints regarding the maximum tolerable fluid flow may indicate that the moderate-deep assemblage is robust to turbulent water flow (Figure~\ref{fig:4p-3d}). Moreover, this   may lead us to investigate factors other than the flow velocity that prevent the assemblage from colonising in shallow water environments. Therefore, the results provide insight into the influence of the flow velocity on coralgal assemblage accretion.  

Similarly, we infer from log-likelihood surface that $\varepsilon$ has a marginally greater control on the population dynamics of assemblages than $\alpha_s$ (Figure~\ref{fig:2p-3d-t}). The Malthusian parameter ($\varepsilon$) governs the intrinsic rate of growth and decline of coral populations, whereas the sub-diagonal assemblage interaction matrix parameter ($\alpha_s$) governs the intensity of the competition between coralgal assemblages. The inference is that the competition between assemblages is less important than the rate of population growth in determining biological interactions between assemblages. 

In summary, \textit{Bayesreef} has enabled greater insight into the importance of the Malthusian parameter and assemblage interaction matrix parameters. Furthermore, it is demonstrably useful in quantifying unobservable parameters such as a coralgal assemblage's response function to long-term hydrodynamic energy exposure.

\subsection{Non-unique solutions }

 The  results show that reef-core predictions are better using a time rather than a depth based likelihood. This is best highlighted by the log-likelihood of $\alpha_s$ and $\varepsilon$ (Figure~\ref{fig:2p-3d}). We observe only a narrow peak in  maximum likelihood using time-based likelihood, and a large plateau of maximum likelihood using depth-based likelihood (Figure~\ref{fig:4p-3d}). The flat surface of the depth-based likelihood indicates  a variety of combinations of population dynamics parameters can give similar observed stratigraphy. Therefore, the depth-based likelihood of \textit{pyReef-Core} simulations can be considered to have no unique solution.

In some cases, we observed that \textit{Bayesreef} cannot reliably estimate the true parameter values, while at the same time produce accurate prediction of the drilled cores which implies multi-modality which is also visible by the likelihood surface. Such cases of multi-modality also appear in other geoscientific problems such as landscape evolution models where different combinations of parameters such as precipitation and landscape erodibility have shown to produce similar topography evolution  \cite{chandra2018bayeslands}. The non-uniqueness in solutions  makes it difficult to disentangle the different process parameters (i.e. environmental parameters and population dynamics parameters) that produced a particular stratigraphy and consequently several competing hypothesis regarding the dominant controls on reef development \cite{Dechniketal16,barrett2017}.   Reef geologists get a limited understanding of the temporal evolution of reefs based on the composition and depth-based likelihood of a drill core alone. The drill cores must have core samples radiometrically dated to be able to constrain the timing of reef accretion, the rates of coralgal accumulation and the rates of sedimentation.  

With \textit{Bayesreef}, we can constrain the environmental conditions that elicit growth responses from corals over time, and thus can better understand the dominant controls on reef development and compare competing hypotheses of reef evolution (e.g.~\cite{NeumannMacintyre85,barrett2012holocene}). In this way, we can isolate the far-fewer parameter combinations that produce a particular time series of reef-growth events and hiatuses. Although \textit{pyReef-Core} a useful tool for reef geologists who wish to understand a reef's temporal evolution,  \textit{Bayesreef} is a significant contribution as it can constrain environmental and biological controls using a core's time structure. To our knowledge, there are no other tools available for studying temporal reef evolution in a statistically robust way.

\subsection{Implications and extensions }

 In future experiments,  we can implement  parallel tempering  MCMC  using a multi-core architecture for high performance computing in order to address the computational requirements when the number of parameters increases. Such implementations have shown to be very useful for inference of landscape evolution models  \cite{chandra2018PT-Bayes}.

The use of a multinomial likelihood produces accurate predictions, however one can argue that it does not fully convey the nature of reef-building processes since it  assumes  independence of observations. In other words, deposition of an assemblage in the past is independent of deposition in the present. In reality, the substrate composition of a reef is a strong predictor of which assemblage will colonise it since corals tend to grow vertically on their skeletons and populate areas where assemblages of their type flourish \cite{Babcock1991}. A degree of dependence between observations is captured by \textit{pyReef-Core}, where the assemblage 'populations' depend on the relative abundance of their same assemblages. Nevertheless, future developments of \textit{Bayesreef} must account for the dependence of observations through time to better capture the nature of biological reef-building processes.

  \textcolor{black}{We  can facilitate other proposal distributions and  MCMC methods in  \textit{Bayesreef}.  Unfortunately more efficient proposals such as those that use  Langevin-gradient MCMC, or the No-U-Turn-Sampler (NUTS),  \cite{Homan2014}  cannot be used, because gradient information is not available from {\it py Reef-core}. However, we can take advantage of the use of heuristic methods for efficient proposal distribution that feature constraint satisfaction  \cite{kumar1992algorithms}. We can use other adaptive MCMC methods such as   \textit{delayed rejection adaptive metropolis, and delayed rejection evolution adaptive metropolis} \cite{vrugt2009accelerating, haario2006dram}  provided that they can  suit the unique constraints in the proposals required for \textit{pyReef-Core}.  }

\section{ Conclusion and Future Work }

We   presented \textit{Bayesreef}, which is a comprehensive Bayesian framework that incorporates multiple sources of information, including forward models, priors and empirical data from geological reef-cores. The Bayesreef framework employed single-chain and parallel  tempering MCMC sampling to address the challenges of multimodal posteriors distributions given synthetic reef-core  and an application from the Great Barrier Reef.   The framework addresses geophysical inverse problem posed by unobserved environmental conditions  and non-unique pathways to reef stratigraphies. The results show that the framework efficiently estimates and provides uncertainty quantification of the  parameters that represent environment and ecological conditions in \textit{pyReef-Core} using the  (posterior) probability distribution.

  In future work, we plan to \textcolor{black}{further enhance  parallel tempering MCMC using better proposal distributions and implementation in high performance computing to address the computational requirements when the number of parameters and the depth  of the reef-core increases.}  textcolor{black}{Furthermore, we can increase the number of assemblages  to give a more realistic view of coralgal evolution in reef-systems; however, this further increases the complexity of the problem. }

\section*{Acknowledgements} 
We would like to 
acknowledge Australian Research Council  (ARC - DP120101793 and FT140101266) and 
 Sydney Research Excellence Initiative 2017, The University of Sydney   for providing the grants for supporting this work. We also wish to acknowledge Sydney Informatics Hub for providing research engineering support for this project. 

\section*{Data and software}

The source code that implements the framework of { \it Bayesreef} can be downloaded from the   Github repository \footnote{https://github.com/intelligentEarth/Bayesreef/tree/master/MCMC\_Sampling}. 

Supplementary results for selected experiments are given here:
\footnote{https://github.com/intelligentEarth/Bayesreef/tree/master/MCMC\_Sampling/supplementary\_results}

\section{References}

\bibliographystyle{IEEEtran} 
\bibliography{bibliography,sample}

\end{document}